\documentclass[prd,eqsecnum,floats,nopacs]{revtex4}
\usepackage{amsmath}
\usepackage{amssymb}
\usepackage{amsfonts}
\usepackage{amsxtra}

\newcommand {\bb}[1]{\mbox{\boldmath $#1$}}

\begin{document}

\title {Pure quantum states of a neutrino with rotating spin in dense magnetized matter}

\author{E. V. Arbuzova}\email{arbuzova@uni-dubna.ru}
\affiliation {International University ``Dubna'', 141980 Dubna,
Russia}
\author{A. E. Lobanov}\email{lobanov@phys.msu.ru}
\affiliation {Department of Theoretical Physics, Moscow State
University, 119991 Moscow, Russia}
\author{E. M. Murchikova}\email{murchikova@gmail.com}
\affiliation {Skobeltsyn Institute of Nuclear Physics, Moscow State University,
119991 Moscow, Russia}

\begin{abstract}

The problem of neutrino spin rotation in dense matter and in strong electromagnetic
fields is solved in accordance with the basic principles of
quantum mechanics. We obtain a complete system of wave functions for a
massive Dirac neutrino with an anomalous magnetic moment which are the eigenfunctions of the kinetic momentum operator and  have the form of
nonspreading wave packets. These wave functions enable one to consider the states
of neutrino with rotating spin as pure quantum states and can be used for calculating  probabilities of various processes with the
neutrino in the framework of the Furry picture.

\end{abstract}


\maketitle

\section*{INTRODUCTION}

Neutrino physics is one of the most rapidly developing areas of
high-energy physics. The fundamental experimental result was
obtained in this field in recent years --- neutrino oscillations
were discovered \cite{osc.exp.}. From the theory of this phenomenon
(see, for example, \cite{books}), which is based on the ideas of
Pontecorvo \cite{Pontecorvo} and Maki, Nakagawa and Sakata
\cite{MNS}, it follows that oscillations are possible only when the
neutrino mass is nonzero. As a consequence of this circumstance, the
possibility exists of neutrino spin rotation, i.e. of transitions
between the active left-handed neutrino polarization state and the
sterile right-handed state. As it is understood today, in contrast
to flavor oscillations, this process can never be observed if the
particle moves in vacuum. For the effect of spin rotation to exist
some external influence leading to effective breaking of the
Lorentz-symmetry of the theory is needed. For instance, dense matter
or electromagnetic fields can serve as the physical reason for this
phenomenon.
In the latter case, nontrivial electromagnetic
properties of the neutrino, in particular, the anomalous magnetic
moment of a Dirac neutrino \cite{LeeShr77}, can lead to rotation of
its spin.

The conventional approach \cite{theor}, \cite{nun} to the theory of
neutrino spin rotation, based on solving the Cauchy problem for
the Schr\"{o}dinger-type equation with an effective Hamiltonian,
was developed for the description of ultrarelativistic
neutrinos (the particles that are observed in experiments now
available). This approach is in fact quasiclassical and is equivalent
to solving the Bargmann--Michel--Telegdi (BMT)
equation \cite{BMT} in the spinor representation (see
\cite{16}).

The aim of this work is to develop a consistent quantum theory of
neutrino spin rotation to describe
neutrinos with nonzero mass, including the low-energy region that may play a significant role in
astrophysics \cite{Fargion}.
Following papers \cite{theor}, \cite{nun},
we use the method of effective potential \cite{Wolf}
that allows us to take into account
collective influence of background particles on the propagation of
neutrino.

To this end, we find a complete system of solutions of the
Dirac-type equation
that describe the action of both electromagnetic field and dense
matter on neutrino dynamics. These solutions have the form of
nonspreading wave packets. They enable one to consider the states
of neutrino with rotating spin as pure quantum states and to
evaluate probabilities of various processes with  participation of
a neutrino in the framework of the Furry picture \cite{F51}.

The paper is organized as follows. In the first Section we present
the model we are working with and briefly describe how the effective
potentials are constructed. In Section \ref{II} we summarize the main
ideas of the method used for solving the Dirac equation. In Section
\ref{III} we introduce a kinetic momentum operator and justify the
advantages of using its eigenvalues as quantum numbers, instead of
the eigenvalues of the canonical momentum operator. Then we find
both stationary and nonstationary wave functions. Here we also derive
the dispersion law and calculate the group velocity for a neutrino
in the dense magnetized matter. We demonstrate that the group velocity
of a neutrino is independent of its spin orientation.
In Section \ref{IV} we find the
explicit expressions for the kinetic momentum and spin projection
operators. In Section \ref{VV} we use the obtained results for
quasiclassical interpretation of neutrino spin behavior in dense
magnetized matter. In the conclusive Section \ref{VI} we compare our
results with those obtained by other authors.

The units $\hbar =c=1$ are used throughout the paper; metric tensor, $g^{\mu\nu}={\mathrm{diag}(1,-1,-1,-1)}$; totally antisymmetric Levi-Civit\`{a} tensor,
$e^{\mu\nu\rho\lambda}= +1$ if $\{\mu\nu\rho\lambda\}$ is an even permutation of $\{0,1,2, 3\}$, $e^{\mu\nu\rho\lambda}= -1$
if it is an odd permutation, $e^{\mu\nu\rho\lambda}= 0$ otherwise.

\section{Wave equation for neutrino in dense matter and in electromagnetic
field}\label{I}

Let us first discuss the model we use for the
description of the neutrino behavior in the matter and in the
electromagnetic field.
At present, there are no experimental
data that might help to make the choice between
the  Dirac or the Majorana nature of a neutrino \cite{pdg}. Therefore, in this paper we discuss spin rotation of  a massive Dirac neutrino with anomalous magnetic moment
propagating through dense magnetized matter.

As a first approximation, we can consider spin rotation
independently of variations of the particle flavor, i.e. we can deal
with the mass eigenstates of neutrino. It is quite obvious, due to the fact that
the physical origins of neutrino flavor
oscillations and its spin rotation are different. The Dirac--Pauli equation \cite{Pauli}
serves as the basis of this approach in electromagnetic fields.
Just in this way, the behavior of the neutrino spin in a constant
homogeneous magnetic field was investigated in \cite{Fu}.
In order to consider the influence of matter on neutrino dynamics through weak interaction,
the Dirac--Pauli equation should be modified.

As it was proposed in \cite{Wolf}, \cite{Mik}, if matter
density is high enough for considering weak interaction of neutrino
with the background fermions as coherent, it is possible to describe
neutrino interaction with matter by an effective potential.
To get covariant realization of this idea, it
is necessary to take into account that the term that describes
direct interaction of neutrino with matter must depend on the
currents
\begin{equation}\label{3}
j_{f}^\mu=\{n_f u^{0}_f,n_f{\bf{u}}_f\},
\end{equation}
and on the polarizations
\begin{equation}\label{4}\lambda^{\mu}_f =\left\{n_f
({\bb{\zeta}}_f\cdot{\bf{u}}_f), n_f\left({\bb{\zeta}}_f + \frac{
{\bf{u}}_f ({\bb{\zeta}}_f\cdot{\bf{u}}_f)}{1+u^{0}}\right)\right\}
\end{equation}
\noindent of the background fermions $f$ \cite{RT,StT,SpinLight}. It
is significant that only $j_{f}^\mu$ and $\lambda_{f}^\mu$ characterize matter
integrally.
In formulas (\ref{3}) and (\ref{4}), $n_f$ and
${\bb{\zeta}}_f \; (0\leqslant |{\bb{\zeta}}_f |^2 \leqslant 1)$ are
the number density and the mean value of the polarization vector of
the background fermions $f$ in the center-of-mass system of matter
($u^{\mu}_f =\{u^{0}_{f},{\bf{u}}_{f}\}$ denotes the four-velocity
of the reference frame), where the mean momentum of the fermions $f$
is equal to zero.

In the framework of the standard model (for a more general case, see
Appendix \ref{uuur}), the interaction term can be derived directly
from the model Lagrangian. To this end, it is necessary to write
this Lagrangian in the low-energy effective form  using the
four-fermion interaction and replace the vector
${\bar{\psi}}_{f}(x)\gamma^{\mu}{\psi}_{f}(x)$ and axial vector
${\bar{\psi}}_{f}(x)\gamma^{5}\gamma^{\mu}{\psi}_{f}(x)$ currents of
the background fermions by $j^{\mu}_{f}$ and $\lambda^{\mu}_{f}$
respectively. The explicit form of $j^{\mu}_{f}$ and
$\lambda^{\mu}_{f}$ is obtained as a result of averaging of currents
over the fermion statistical distribution function and depends on
the type of the  matter considered. In particular, the expression
for the polarization $\lambda^{\mu}_{\mathrm{e}}$ of electron gas in
a magnetic field at finite temperature can be found in \cite{nun}.
Note that, since the origin of the effective potential is forward
elastic scattering of neutrino, it can also be calculated using
field-theoretic methods \cite{160}.

As a result we come to the equation
\begin{equation}\label{6}
\left(i{\gamma^{\mu}\partial_{\mu}}  - \frac{1}{2}
\gamma^{\mu}{f_{\mu}}(1 + \gamma ^5) - \frac{i}{2}\mu_{0}F^{\mu \nu
}\sigma _{\mu \nu } - m\right)\varPsi(x) = 0,
\end{equation}
\noindent where $F^{\mu\nu}$ is
the electromagnetic field tensor. The $\gamma$-matrices satisfy
$\gamma^{\mu}\gamma^{\nu}+\gamma^{\nu}\gamma^{\mu}=2g^{\mu\nu}$
with $\gamma^{0}$ Hermitian, ${\bb{\gamma}}$  anti-Hermitian, and are related to the $\beta$ and ${\bb{\alpha}}$ matrices through
$\gamma^{0}=\beta, {\bb{\gamma}}=\beta{\bb{\alpha}}$;
$\sigma ^{\mu \nu }
=\frac{1}{2}(\gamma^{\mu}\gamma^{\nu}-\gamma^{\nu}\gamma^{\mu})$,
$\gamma^{5}=-i\gamma^{0}\gamma^{1}\gamma^{2}\gamma^{3}$. The effective
four-potential $f^\mu$ is a linear combination of currents and
polarizations of the background fermions
\begin{equation}\label{2}
f^\mu =
{\sum\limits_f\big({\rho_f^{(1)}j_f^\mu+\rho_f^{(2)}\lambda_f^\mu}\big)}.
\end{equation}
\noindent Summation in (\ref{2}) is carried out over all fermions $f$ of
matter. The expressions for the coefficients $\rho_{f}^{(1,2)}$
are determined as
\begin{equation}\label{rho}
\rho_f^{(1)}\!=\sqrt 2 {G}_{{\mathrm F}} \left\{
I_{\mathrm{e\nu}}+T_3^{(f)}-2Q^{(f)}\sin^2\theta_{\mathrm W} \right\}\!,\;\;
\rho_f^{(2)}\!=-\sqrt 2 {G}_{{\mathrm F}} \left\{ I_{\mathrm{e\nu}}+T_3^{(f)}
\right\}\!.
\end{equation}

\noindent Here $Q^{(f)}$ is the electric charge of the fermion $f$;
$T_{3}^{(f)}$ is the third component of the weak isospin; ${G}_{{\mathrm
F}}$ and $\theta_{\mathrm W}$ are the Fermi constant and the
Weinberg angle respectively; \mbox{$I_{\mathrm{e\nu}}=1$} for the electron
neutrino interaction with electrons, \mbox{$I_{\mathrm{e\nu}}=-1$} for
interaction with positrons, otherwise $I_{\mathrm{e\nu}}= 0$.

While equation (\ref{6}) describes mass states of neutrino,
potential (\ref{2}) is flavor dependent, and in the general case, this
leads to correlations between spin rotation and flavor oscillations.
In order to construct a mathematically consistent approach to
a description of neutrino spin rotation, we have to avoid the
correlations. This is possible if we assume that the effective potential
which describes the influence of dense matter on neutrino is the
same for different flavors. In this case we can construct flavor
states of the neutrino as the linear combination of its mass states with
coefficients that are elements of the mixing matrix of the neutrino in
vacuum.

The physical justification of this model can be made in the
following way. In order to consider effective potentials as
flavor independent, it is necessary to assume
that the fraction of electrons in the matter is small, i.e. the
electron number density ${n_{e}}$ is approximately equal to zero.
Calculations which are in a satisfactory agreement with the
experimental data show that in the center of a neutron star the
fraction of electrons does not exceed a few per cent \cite{stars}.
Therefore the model approximation $ {n_{e}= 0}$ is quite
appropriate.

We restrict ourselves to considering equation (\ref{6}) in the case
of a constant homogeneous electromagnetic field and constant
currents and polarizations of matter,
\begin{equation}\label{5.0}
    F^{\mu\nu}={\mathrm{const}}, \quad j_f^{\mu}={\mathrm{const}}, \quad
    \lambda_f^{\mu}={\mathrm{const}},
\end{equation}
\noindent as the first order approximation of the realistic
background.
Because of (\ref{5.0}) additional conditions for
$F^{\mu\nu}$ and $f^{\mu}$ may be obtained. The strengths of the
electric and magnetic fields and average currents and polarizations
of the background particles should obey the self-consistent system
of equations including the Maxwell equations, the Lorentz equation
\begin{equation}\label{01}
    {\dot{j}}^\mu_{f}=\frac{e_{f}}{m_{f}}
    F^{{\mu}{\phantom{\nu}}}_{{\phantom{\mu}}{\nu}}j^\nu_{f},
\end{equation}
\noindent and the the BMT quasiclassical
spin evolution equation
\begin{equation}\label{02}
   {\dot{\lambda}}^\mu_{f}=\left[\frac{e_{f}}{m_{f}}
    F^{{\mu}{\phantom{\nu}}}_{{\phantom{\mu}}{\nu}}
    +2\mu_{f}
    \left( g^{{\mu}{\alpha}}
    -u^{\mu}_{f}u^{\alpha}_{f}\right)F_{\alpha\nu}
    \right]\lambda^\nu_{f}.
\end{equation}
\noindent Here the dot denotes differentiation with respect to the
proper times $\tau_{f}$ of particles. From equations (\ref{01}) and
(\ref{02}) we find that for  charged particles the conditions
$\dot{j}^\mu_{f}=0,\; \dot{\lambda}^\mu _{f}=0$ are equivalent to
$F^{{\mu}{\phantom{\nu}}}_{{\phantom{\mu}}{\nu}}j^\nu_{f}=0,$
$F^{{\mu}{\phantom{\nu}}}_{{\phantom{\mu}}{\nu}}\lambda^\nu_{f}=0$.
However, it seems reasonable to assume that in dense matter
velocities of
the center-of-mass systems
for all components are
equal, thus for the neutral particles similar conditions should
hold as well. In this way we
obtain the restriction
\begin{equation}\label{03}
    F^{\mu\nu}f_\nu=0.
\end{equation}
\noindent It should be emphasized
that condition (\ref{03}) is the result of the
fact that average currents and polarizations of particles of
matter in an external field should satisfy the classical equations
of motion.  The physical
meaning of this condition is discussed in more
detail in Section \ref{VV}.

Note that equation (\ref{6}) with constant coefficients appears in
the standard model extension \cite{2}, \cite{nambu} as well. In this case,
$F^{{\mu}{\nu}}$ and $f^{{\mu}}$ describe vacuum condensates that break the Lorentz-symmetry of the theory.

\section{Formulation of the problem}\label{II}

The Dirac equation, in particular equation (\ref{6}), is a
partial differential equation. Therefore, as it is known, its
general solution is defined up to an arbitrary function.
However, in quantum-mechanical
applications we deal with the so-called complete integral which depends
on a set of constants, i.e. quantum numbers. Those are the eigenvalues
of some self-adjoint operators. For the classification of
particle states it is necessary to introduce the complete set of the
operators --- integrals of motion. Note that in nonrelativistic
quantum mechanics any self-adjoint operator may serve as an
operator of observable, which is not the case in the relativistic mechanics,
where only integrals of motion, commuting with the operator of the
equation, can be treated as operators of observables \cite{LanP}.
The choice of the complete set of them
is different in each particular case and should be adequate to the
problem being solved.

In the case of the problem of spin evolution, operators of kinetic momentum components should be included in the complete set. It becomes obvious if we start with the
following argument. The direction of particle polarization is well defined in
its rest frame, and then one finds the polarization in the
laboratory frame upon carrying out an appropriate Lorentz transformation. This transformation is
defined by the group velocity of the particle, in other words,
by its kinetic momentum. This implies that it is not the canonical momentum but the kinetic momentum operator that defines the direction of particle propagation.

Now, there is only one problem that remains to be solved. The form of the
kinetic momentum operator for a
particle with spin propagating under the influence of external fields is not known beforehand. So we
have to find a self-adjoint operator ${\mathfrak{p}}^{\mu}$ with
the eigenvalues $q^{\mu}$, which satisfies the condition $q^{2}=m^{2}$
and may be interpreted as the components of the particle kinetic momentum.

Let us discuss this issue in more detail.
In the mathematical apparatus of quantum field theory, a particle is
usually identified with an irreducible unitary representation of the
Poincar\'{e} group \cite{BoLoTO}. The irreducible representations are
characterized by two invariants of the group
\begin{equation}\label{a4}
P^2 \equiv{P^{\mu}}{P_{\mu}}=m^2,
\end{equation}
\begin{equation}\label{a5}
W^2 \equiv {W^{\mu}}{W_{\mu}} = -{m^2}s(s+1).
\end{equation}
\noindent The translation generators $P^\mu$ are identified with the
components of the particle momentum, and the Pauli--Lubanski--Bargmann vector
\begin{equation}\label{a6} W^{\mu}
  =-\,{\frac{1}{2}}\,{e^{{\mu}{\nu}{\rho}{\lambda}}}{M_{{\nu}{\rho}}}
  {P_{\lambda}},
\end{equation}
\noindent where $M^{{\mu}{\nu}}$ are the Lorentz generators,
characterizes the particle spin. The invariant $m^2$ is the particle
mass squared and $s$ is the value of its spin.

A space of unitary representation is defined by the condition called
``the wave equation for a particle with mass $m$ and spin $s$.'' The
wave equation for particles with spin $s ={1}/{2}$ is the Dirac
equation
\begin{equation}\label{a9}
\left(i\gamma^{\mu}\partial_{\mu} - m\right){\varPsi}_{0}(x) = 0.
\end{equation}
\noindent In this case the realization of generators of the Poincar\'{e}
group and the Pauli--Lubanski--Bargmann vector in the coordinate
representation is
\begin{equation}\label{a10}
  p^{\mu} = i{\partial^{\,\mu}},\quad m^{\mu\nu}=
  i(x^{\mu}\partial^{\nu}-x^{\nu}\partial^{\mu})+\frac{i}{2}\sigma^{\mu\nu},\quad
  w^\mu =
  \frac{i}{4}\,{\gamma^{5}}\sigma^{\mu\nu}\partial_{\nu}.
\end{equation}

These operators commute with the operator of the Dirac equation and
can be identified with observables.  They have a
self-adjoint extension on the subsets of solutions of equation
(\ref{a9}) with a fixed sign of the energy with regard to the standard scalar product,
\begin{equation}\label{a010}
({\varPhi}, \varPsi)= \int\! d{{\bf
x}}\,{\varPhi}^{\dag}({\bf x},t)\,\varPsi({\bf x},t).
\end{equation}
\noindent Three-dimensional spin projection operator
$\bb{{s}}(p)$ is a set of coefficients
${{s}}_{i}(p)$ of the expansion of the vector $w^\mu$ in
spacelike unit vectors ${{S}}_{i}^{\mu}(p)\; (i=1,2,3)$:
\begin{equation}\label{a11}
  s_{i}(p) = -\,\frac{1}{m}w_{\mu}S_{i}^{\mu}(p),
\end{equation}
\noindent where
\begin{equation}\label{a110}
p_{\mu}S^{\mu}_{i}(p) =0,\quad S^{\mu}_{i}(p)S_{\mu
j}(p)=-\delta_{ij}.
\end{equation}
\noindent Obviously,
\begin{equation}\label{a12}
\big[s_{i}(p),s_{j}(p)\big]
=i{e_{ijk}}s_{k}(p).
\end{equation}
\noindent The choice of these unit vectors is not unique, and it is
possible to construct operators that determine the spin projection
on any direction in an arbitrary Lorentz frame.

The above description of the particle characteristics cannot be
directly used in the presence of external fields, where operators (\ref{a10}) are not necessarily integrals of
motion. In this case the classification of particle states is usually realized  by linear combinations of operators
$p^{\mu}$ and $w^{\mu}$ with coefficients depending on coordinates
\cite{BG90}.
Unfortunately, the physical meaning of these operators  is usually not quite clear.
Therefore, one should formulate a description of the particle motion in an external field in the same clear and detailed way as for a free particle.

This can be achieved basing on the following ideas.
Since an irreducible representation of group is defined accurately
up to an equivalence transformation, it is reasonable to state the
problem of finding such realization of the Lie algebra of the
Poincar\'{e} group for which the condition of
irreducibility of the representation leads to wave equation describing a particle in a given external background. To solve this problem it is necessary to
find a unitary operator $U(x,x_{0})$ which converts solutions of the
wave equation for a free particle (\ref{a9}) to solutions
$\varPsi(x)$ of (\ref{6}):
\begin{equation}\label{a018}
U(x,x_{0})\varPsi_{0}(x)=\varPsi(x).
\end{equation}
\noindent  Thus, $U(x,x_{0})$ is an intertwining operator in the
sense of Darboux \cite{Darboux}.  This operator in our case should
satisfy the equation
\begin{equation}\label{a19}
 \left(i\gamma^{\mu}{\partial}_{\mu} - \frac{1}{2}
\gamma^{\mu}{f_{\mu}}(1 + \gamma ^5) - \frac{i}{2}\mu_{0}F^{\mu \nu
}\sigma _{\mu \nu }\right)U(x,x_{0}) -
U(x,x_{0})\left(i{\gamma^{\mu}\partial_{\mu}}\right) =0.
\end{equation}
\noindent Therefore, operators
\begin{equation}\label{a21}
  {\mathfrak{p}}^{\mu} =U(x,x_{0})p^{\mu}U^{-1}(x,x_{0}),\quad {\mathfrak m}^{\mu\nu}
  =U(x,x_{0})m^{\mu\nu}U^{-1}(x,x_{0})
\end{equation}
\noindent commute with the operator of the wave equation. As a
consequence, the Pauli--Lubanski--Bargmann vector ${\mathfrak{W}}^{\mu}$ and the components
of the three-dimensional spin projection operator ${\mathfrak{S}}_{i}$
can be constructed in the same way as in the case of a free particle:
\begin{equation}\label{aa21}
  {\mathfrak{W}}^{\mu}
  =-\,{\frac{1}{2}}\,{e^{{\mu}{\nu}{\rho}{\lambda}}}{{\mathfrak{m}}_{{\nu}{\rho}}}
  {{\mathfrak{p}}_{\lambda}}, \quad {\mathfrak{S}}_{i}=
  -\,\frac{1}{m}{\mathfrak{W}}_{\mu}S_{i}^{\mu}({\mathfrak{p}}).
\end{equation}

The above statement may be reduced to the following: the wave
function of a neutrino in dense matter can be derived with the help
of a solution of the Dirac equation for a free particle and of some
unitary evolution operator. A complete set of integrals of motion
may be constructed with the help of operators (\ref{a21}). The
physical meaning of eigenvalues of observables, i.e. quantum
numbers, is clear enough then. However,
in the general case, $U(x,x_{0})$ is
an integral operator and, as a consequence, operators (\ref{a21})
are also integral ones, so it is difficult to find
$U(x,x_{0})$ by direct calculations.
That is why we are trying to find the evolution operator for equation
(\ref{6}) using the correspondence principle.

\section{Wave functions}\label{III}

When motion of a massive particle with spin in electromagnetic fields is
described in the framework of the quasiclassical approach, we
use two four-vectors, namely the four-velocity $u^{\mu}$ and the spin
vector $S^{\mu}$ obeying the conditions
\begin{equation}\label{x1}
u^{2}=1,\qquad S^{2}=-1,\qquad (uS)\equiv u^{\mu}S_{\mu}=0.
\end{equation}
\noindent  These vectors are solutions of the Lorentz
and BMT equations, respectively. As it was shown in \cite{Zw},
both four-vectors can be constructed with the help of one and the same evolution operator that acts on different initial values of these four-vectors, satisfying
relations (\ref{x1}).
In other words, the evolution operator for the BMT equation completely describes
quasiclassical behavior of the particle.

As far as the neutrino is a neutral particle,  the BMT equation for the spin
vector $S^{\mu}$ of a neutrino moving with the four-velocity $u^{\mu}$
has the form
\begin{equation}\label{BMT1}
    \dot{S}^{\mu}=2\mu_{0}\left(g^{\mu\alpha}- u^{\mu}u^{\alpha}\right)F_{\alpha\nu} S^{\nu} .
\end{equation}
\noindent
We can extend the BMT equation to include weak interaction
between the neutrino and dense matter. It was shown phenomenologically
in \cite{L43} that if effects of neutrino weak interactions are
taken into account, the Lorentz-invariant generalization of the BMT
equation is
\begin{equation}\label{x4}
 \dot{S}^{\mu}=2 \left(g^{\mu\alpha}- u^{\mu}u^{\alpha}\right)\left(\mu_{0} F_{\alpha\nu}+G_{\alpha\nu} \right) S^{\nu},
\end{equation}
\noindent where
\begin{equation}\label{5}
G^{\mu \nu}= \frac{1}{2}e^{\mu \nu \rho \lambda}
f_{\rho}u_{\lambda}.
\end{equation}
\noindent We can get an analogous result by averaging the equations of
motion for the Heisenberg operators (see \cite{FRG}).

A remarkable feature of the BMT equation is its universality --- at least in the zeroth order of
the Planck constant, particles with an arbitrary spin may be
described by the same BMT equation. However, a part of the information
about the particle behavior gets lost due to distinctions in
transformation properties of the states of the particles with
different spins. This information may be restored in the following
way. We can describe a particle using quasiclassical spin
wave functions $\varPsi(\tau)$ constructed in such a way that the
vector current
\begin{equation}\label{V}
j _{V}^{\mu}(\tau)=\bar{\varPsi}(\tau)\gamma^{\mu}\varPsi(\tau),
\end{equation}
\noindent as well as the axial current
\begin{equation}\label{A}
j_{A}^{\mu}(\tau)=\bar{\varPsi}(\tau)\gamma^{5}\gamma^{\mu}\varPsi(\tau)
\end{equation}
\noindent built on their base, obey  the BMT equation, and, as a
consequence, may be interpreted as four-velocity $u^{\mu}$ and spin
vector $S^{\mu}$ respectively.

We can introduce quasiclassical spin wave functions as follows
\cite{LP}, \cite{L}. Let the Lorentz equation be solved, i.e. the
dependence of the particle coordinates on proper time is found.
Then the BMT equation transforms to an ordinary differential equation,
whose resolvent determines a one-parameter subgroup of the
Lorentz group. The quasiclassical spin wave function is a
spin tensor, whose evolution is determined by the same
one-parameter subgroup. It is easy to verify (see Appendix \ref{rr}) that for a neutral particle, represented by a
Dirac bispinor, the equation that defines the evolution operator for
the wave function $\varPsi(\tau)$ takes the form
\begin{equation}\label{103}
\dot{U}(\tau,\tau_{0}) = \frac{i}{m}\gamma^{5}\sigma_{\mu\nu}
{\varphi}^{\mu}{q}^{\nu}\,
{U}(\tau,\tau_{0}).
\end{equation}
\noindent Here $\varphi ^\mu = f^\mu/2 + H^{\mu\nu}q_{\nu}/m$, $H^{\mu\nu}=\mu_{0}{}^{\star\!}F^{\mu\nu}$, where ${}^{\star\!}F^{\mu\nu} =
-\frac{1}{2}e^{\mu\nu\rho\lambda}F_{\rho\lambda}$ is the dual
electromagnetic field tensor and $q^{\mu}=mu^{\mu}$.

Let us look for a solution of equation (\ref{6}) in the form
\begin{equation}\label{102}
\varPsi (x) = e^{-iK(x)}U\left(\tau (x),0\right)\varPsi_{0}
(x),
\end{equation}
\noindent where $U(\tau,\tau_{0})$ is an evolution operator for the
quasiclassical spin wave function,  $e^{-iK(x)}$ is a phase factor, and
$\varPsi _0 (x)$ is a solution of the Dirac equation for a free
particle.
Since we assume that $F^{\mu\nu}$ and
$f^{\mu}$ are constants, the evolution operator can be expressed
as a matrix exponential,
\begin{equation}\label{U}
    U(\tau,\tau_{0}) = \exp \left\{ \frac{i}{m}{
    \gamma^{5}\sigma_{\mu\nu}
{\varphi}^{\mu}{q}^{\nu}
    (\tau-\tau_{0})} \right\},
\end{equation}
\noindent that depends on constant four-vector $q^\mu$ satisfying
the condition $q^2=m^2$.

Using ansatz (\ref {102}) we should choose such a basis in the space of
solutions of the Dirac equation for the free particle that the action
of the evolution operator  on each element of this basis be reduced
to multiplication by one and the same matrix function depending on quantum
numbers of the basis. It is obvious that in our case we must choose
the plane waves
\begin{equation}\label{a3}
\varPsi _0 (x) = e^{ - i(qx)}(1 - \zeta_{0}\gamma
^5\gamma_{\mu}{S}_0^{\mu}(q) )(\gamma^{\mu}q_{\mu} + m)\psi _0.
\end{equation}
\noindent Here ${S}_0^{\mu}(q)$ determines the direction of
polarization of the particle; $\zeta_{0} = \pm 1 $ is the spin
projection on $ {S}_0^{\mu}(q)$; $\psi _{0}$ is a constant
four-component spinor. The wave function is normalized by the condition
\begin{equation}\label{aaa3}
 {\bar{\varPsi}}_{0}(x)\varPsi_{0}(x) =m/q_{0},
\end{equation}
\noindent and four-vector $q^\mu$ is a kinetic momentum of the particle. Though the explicit form of a kinetic momentum
operator for a particle with spin interacting with dense matter and
electromagnetic field is not known beforehand, the correspondence
principle allows us to construct solutions characterized by its eigenvalues.

Naturally, as far as equation (\ref {6}) is invariant under
translations the canonical momentum operator
$p^{\mu}=i{{\partial}^{\mu}}$ is an integral of motion for this
equation, too. However, the commonly adopted choice of eigenvalues
of this operator as quantum numbers is not satisfactory if we prefer
spin projection operators with clear physical meaning. The
directions of canonical and kinetic momenta are different in the general
case (see \cite {nambu}) and, as it
was already mentioned in Section \ref{II}, projection of the spin is well defined in the rest frame of the particle where its kinetic momentum is equal to zero: ${\bf q}=0$.
That is why in the construction of spin projection operators (see
(\ref {a11})) it is necessary to select unit vectors orthogonal to
four-vector $q^{\mu}$.

Let us find the proper time $\tau(x)$ and
the phase factor $K(x)$. Substitution of (\ref{102}) into
(\ref{6}) gives
\begin{equation*}\label{aa6}
\begin{array}{c}
\displaystyle\left\{\gamma^{\mu}{q}_{\mu}
+\gamma^{\mu}{\partial}_{\mu}K(x) -
\frac{1}{2}\gamma^{\mu}{f}_{\mu}(1+\gamma^{5}) + \frac{1}{m}\gamma^{5}\sigma_{\mu\nu}
{\varphi}^{\mu}{q}^{\nu}\gamma^{\alpha}{N}_{\alpha}
\right.\\[8pt] - \displaystyle\left.\frac{i}{2}\mu_{0}F^{\mu \nu }\sigma
_{\mu \nu }-m\right\}e^{-iK(x)}U(\tau(x) )\varPsi_{0} (x)
 = 0,
\end{array}
\end{equation*}
\noindent where $N^{\mu} = \partial ^{\mu} \tau $. Since the matrix
$U(\tau(x))$ is nondegenerate and the commutator $[\gamma^{\mu}q_{\mu},U(\tau(x))]$ is zero, we get
\begin{equation*}\label{a7}
\begin{array}{l}
\displaystyle\gamma^{\mu}{\partial}_{\mu}K(x) -
\frac{1}{2}\gamma^{\mu}{f}_{\mu}(1+\gamma^{5}) + \frac{1}{m}\gamma^{5}\sigma_{\mu\nu}
{\varphi}^{\mu}{q}^{\nu}\gamma^{\alpha}{N}_{\alpha} -
\frac{i}{2}\mu_{0}F^{\mu \nu }\sigma _{\mu \nu }
= 0.
\end{array}
\end{equation*}

To solve this equation for $N^\mu$ and $K(x)$, we should set the
coefficients at the linearly independent elements of the algebra of the Dirac
matrices equal to zero. Thus
\begin{equation}\label{a8}
    {\partial}^{\mu}K(x) = f^{\mu}/2,
\end{equation}
\noindent and the system that defines the vector $N^{\mu}$ is
\begin{equation}\label{106}
\varphi ^\mu (m - (Nq)) + (N\varphi )q^\mu = 0, \quad
     \mu_{0}F^{\mu\alpha}q_{\alpha} = - e^{\mu \nu \rho \lambda }
    N_\nu \varphi _\rho q_\lambda,
\end{equation}

The system is consistent provided that
\begin{equation}\label{dop1}
q_{\mu}F^{\mu\nu}\varphi_{\nu}=0.
\end{equation}
\noindent Equation (\ref{dop1}) must be held for an
arbitrary $q^{\mu}$, so we have
\begin{equation}\label{dop11}
  {}^{\star\!}F^{\mu\alpha}F_{\alpha\nu}\equiv -\frac{1}{4}\delta^{\mu}_{\nu}
  {}^{\star\!}F^{\alpha\beta}F_{\alpha\beta}=0,
\end{equation}
\begin{equation}\label{dop12}
F^{\mu\nu}f_{\nu}=0.
\end{equation}
\noindent Any  antisymmetric tensor has an eigenvector corresponding
to zero eigenvalue if and only if its second invariant
$I_{2}=\frac{1}{4}{}^{\star\!}F^{\mu\nu}F_{\mu\nu}$ is equal to zero. That is
why conditions (\ref{dop11}) and (\ref{dop12}) are not independent
and (\ref{dop11}) is a consequence of (\ref{dop12}). Comparing it with
(\ref{03}), we see that the condition obtained basing on the physical
reasons alone and condition (\ref{dop12}) totally coincide. This ensures quasiclassical behavior of the particle  in the
background medium and allows obtaining the solution of
equation (\ref{6}) in  the form of (\ref{102}).

Using an orthogonal basis in the Minkowski space,
\begin{equation}\label{Basis}
    \displaystyle n_0^\mu=q^\mu/m, \;\;
    \displaystyle n_1^\mu=\frac{H^{\mu\nu}q_\nu}{\sqrt{{\cal N}}},
    \;\;
    \displaystyle n_2^\mu=\frac{F^{\mu\nu}q_\nu}{\sqrt{\tilde{{\cal N}}}},
    \;\;
    \displaystyle n_3^\mu=\frac{m^{2}H^{\mu\nu}H_{\nu\alpha}
    q^\alpha  -q^\mu {\cal N}}
    {m\sqrt{{\cal N}\tilde{{\cal N}}}},
\end{equation}
\noindent where ${\cal N}=q_{\mu}
H^{\mu\rho}H_{\rho\nu}q^{\nu}$, $\tilde{\cal N}=\mu_{0}^{2}q_{\mu}
F^{\mu\rho}F_{\rho\nu}q^{\nu}$, and
relations (\ref{A2.1}), (\ref{A2.3}) from Appendix \ref{ur}, we find
\begin{equation}\label{N}
\displaystyle N^\mu = - q^\mu\frac{m(f\varphi)}
    {{2((\varphi q)^2-m^{2}\varphi^2)}}
     + f^\mu\frac{m}{2(\varphi q)}  + \varphi^\mu\frac{m^3(f \varphi)}
    {2(\varphi q)((\varphi q)^2-m^{2}\varphi^2)}.
\end{equation}
\noindent According to the fact that $f^\mu$ and
$N^\mu$ are constant values  we obtain the proper time
\begin{equation}\label{xx} \tau = (Nx),
\end{equation}
\noindent and the phase factor
\begin{equation}\label{xx0}
K(x) =(fx)/2,
\end{equation}
which determines an energy shift of the neutrino in matter.

Now we can derive the expression for the wave function:
\begin{equation}\label{Wf1}
\displaystyle \varPsi_{q\zeta_{0}}(x) =
\frac{1}{2}\sum\limits_{\zeta = \pm 1} e^{-i(P_{\zeta} x)}{( 1 -
\zeta \gamma^5 \gamma_{\mu}{S}_{\mathrm{tp}}^{\mu}(q))(1-\zeta_{0}\gamma^5
\gamma_{\mu}{S}_{0}^{\mu}(q))(\gamma^{\mu}{q}_{\mu}+m)} \psi_{0},
\end{equation}
where
\begin{equation}\label{P2}
\begin{array}{c}
\displaystyle P^\mu_{\zeta} = q^\mu + f^{\mu}/2-\zeta N^{\mu}\sqrt
    {(\varphi q)^2 - \varphi^2 m^2}/m
\displaystyle = q^\mu \bigg(1+\zeta\frac{(f\varphi)}
    {2\sqrt{(\varphi q)^2-m^2\varphi^2}}
    \bigg)\\[12pt]\displaystyle+ \frac{1}{2}f^{\mu}\bigg(1  - \frac{\zeta
\sqrt{(\varphi q)^2-m^2\varphi^2}}{(\varphi q)}\bigg)\,
    \displaystyle -\,\varphi^\mu \,\frac{\zeta (f \varphi) m^2}
    {2(\varphi q)\sqrt{(\varphi q)^2-m^2\varphi^2}},
\end{array}
\end{equation}
\begin{equation}\label{P3}
\displaystyle {S}_{{\mathrm{tp}}}^\mu(q) = \frac{q^\mu (\varphi q) / m -
    \varphi ^\mu m }{\sqrt {(\varphi q)^2 -
    \varphi ^2m ^2}}.
\end{equation}
\noindent System (\ref{Wf1}) represents the complete system of solutions of equation
(\ref{6}) characterized by kinetic momentum of the particle
$q^{\mu}$ and the quantum number $\zeta_{0}= \pm 1$ which can be
interpreted as the neutrino spin projection on the direction
${S}_{0}^\mu(q)$ at $\tau = (Nx)=0$.

System (\ref{Wf1}) is nonstationary in the general case. The
solutions are stationary only when the initial polarization vector
$S_{0}^{\mu}(q)$ is equal to the vector of the total polarization
$S_{{\mathrm{tp}}}^{\mu}(q)$ \cite{x}, $S_{0}^{\mu}(q)=S_{{\mathrm{tp}}}^{\mu}(q)$. In this case the wave functions are eigenfunctions of the spin
projection operator ${\mathfrak{S}}_{\mathrm{tp}}=-\gamma^5 \gamma_{\mu}{S}_{\mathrm{tp}}^{\mu}(q)$ with eigenvalues $\zeta = \pm 1$, and of the canonical momentum operator $p^{\mu}=i\partial^{\mu}$ with eigenvalues $P^{\mu}_{\zeta}$. The orthonormal system of the stationary solutions, the basis of
solutions of equation (\ref{6}), can be written as (see Appendix \ref{ru})
\begin{equation}\label{Wf}
    \varPsi_{q\zeta}(x) = e^{ - i(P_{\zeta}x)}\sqrt{|J_{\zeta}(q)|}
    (1 - \zeta\gamma^5\gamma_{\mu}{S}_{\mathrm{tp}}^{\mu}(q))(\gamma^{\mu}{q}_{\mu} + m)\psi_0,
\end{equation}
\noindent where $J_{\zeta}(q)$ is the transition Jacobian between the variables
$q^\mu$ and $P^\mu_{\zeta}$
\begin{equation}\label{nor3}
  J_{\zeta}(q)=  {\mathrm{det}}(M_{ij}) =
    {\mathrm{det}} \left[ \frac{\partial P^i_{\zeta}}{\partial
    q^j}+\frac{\partial P^{i}_{\zeta}}{\partial q^{0}} \frac{\partial q^{0}}
    {\partial q^j} \right].
\end{equation}
\noindent With the help of relations from Appendix \ref{ur}, the explicit form of the matrix
$M_{ij}$ may be written as
\begin{equation*}
\begin{array}{c}
    \displaystyle M_{ij}=\delta_{ij}
    \left( 1+\zeta\frac{(f\varphi)}{2\sqrt{(\varphi q)^2-m^2 \varphi^2}}
    \right)\\[20pt]
    \displaystyle+\zeta\left( q_i-\varphi_i\frac{m^2}{(\varphi q)} \right)
    \left( \varphi_j-q_j\frac{\varphi^0}{q^0} \right)\frac{m^2(f \varphi)^2+f^2
    (\varphi q)^2-m^2 f^2\varphi^2}{ 4(\varphi q)((\varphi
    q)^2-m^2\varphi^2)^{3/2}}.
\end{array}
\end{equation*}
\noindent One can easily derive the following equality for arbitrary
vectors ${\bf g}$ and ${\bf h}$
$$\det{(\delta_{ij}+g_{i}h_{j})}=1+({\bf{g}}\cdot{\bf{h}}).$$ So we have
\begin{equation}\label{Det}
    J_{\zeta}(q)={\left({1+\zeta\frac{(f \varphi)}
    {2\sqrt{(\varphi q)^2-m^2 \varphi^2}}}\right)}^2
    {\left({1+\zeta\frac{f_{\mu}H^{\mu\nu}q_{\nu}/(2m)-2\mu_{0}^{2}
    I_1}
    {\sqrt{(\varphi q)^2-m^2 \varphi^2}}}\right)}.
\end{equation}
\noindent Here $I_{1}=\frac{1}{4}F^{\mu\nu} F_{\mu\nu}$ is
the first invariant of the tensor $F^{\mu\nu}$. Note that to obtain
a complete system of solutions for the antineutrino, it is necessary to
change the sign of the kinetic momentum $q^{\mu}$.

The dispersion law for the neutrino in dense magnetized matter is
different from the one for the free particle and can be written as (see
relations (\ref{vvv}), (\ref{vv}) and (\ref{z5}) in Appendix \ref{ur})
\begin{equation}\label{PP3}
    {\Tilde{P}}^2=m^2-f^{2}/4-2\mu_{0}^{2} I_1-2\zeta \Delta\sqrt{{(\Tilde{P}}
    {\Tilde{\Phi}})^2-{\Tilde{\Phi}}^{2}m^2},
\end{equation}
where
\begin{equation}\label{bbb}
\begin{array}{c}
 \displaystyle   \tilde{P}^\mu=P^{\mu}_{\zeta}-f^{\mu}/2, \quad
    \tilde{\Phi}^\mu=f^\mu/2+H^{\mu\nu}\tilde{P}_\nu/{m},\\[8pt]
    \displaystyle\Delta =
{\mathrm{sign}}{\left({1+\zeta\frac{f_{\mu}H^{\mu\nu}q_{\nu}/(2m)-2\mu_{0}^{2}
    I_1}
    {\sqrt{(\varphi q)^2-m^2 \varphi^2}}}\right)}=
    {\mathrm{sign}}\left(1+\frac{f_{\mu}H^{\mu\nu}\tilde{P}_{\nu}/m-4\mu_{0}^{2}
    I_1}{\tilde{P}^2 - m^2 +
f^{2}/4+2\mu_{0}^{2}
    I_1-(\tilde{\Phi}f)}\right).
\end{array}
\end{equation}
\noindent The appearance of the factor $\Delta$ in equation
(\ref{PP3}) is a consequence of the fact that $\zeta$ is projection of the
particle spin on the direction defined by the kinetic
momentum instead of the canonical one.

In spite of the modification of the dispersion law described above, we
see that the neutrino moving through dense matter and
electromagnetic field may still behave as a free particle, i.e.
its group velocity
\begin{equation}\label{gr}
{\bf v}_{{\mathrm{gr}}} = \frac{\partial {P}_{\zeta}^{0}}{\partial {\bf
    P}_{\zeta}}= \frac{{\bf
    q}}{{q}^{0}}
\end{equation}
\noindent is the same for both polarization states of the
particle. However, in interactions with other particles some
channels of reactions which are closed for a free neutrino can be
opened due to the modification of the dispersion law (see, for example, \cite{SpinLight}, \cite{zlm}, \cite{pairs}).

Let us discuss now properties of nonstationary solutions in more
detail. Solution (\ref{Wf1}) is a plane-wave solution of equation (\ref{6}), describing a pure quantum-mechanical state of a neutral particle with a nonconserved spin projection on the fixed space axis. Solutions (\ref{Wf1}) do not form an orthogonal basis. However, the
considered system is not overcomplete, since the spectrum of the spin projection operator is finite. So the system can be easily orthogonalized.
Generalization of the basis (\ref{Wf}) is
\begin{equation}\label{Wf2}
\displaystyle \tilde{\varPsi}_{q\zeta_{0}}(x) =
\frac{1}{2}\sum\limits_{\zeta = \pm 1} e^{-i(P_{\zeta}
x)}\sqrt{|J_{\zeta}(q)|}{( 1 - \zeta \gamma^5
\gamma_{\mu}{S}_{{\mathrm{tp}}}^{\mu}(q))(1-\zeta_{0}\gamma^5
\gamma_{\mu}{S}_{0}^{\mu}(q))(\gamma^{\mu}{q}_{\mu}+m)} \psi_{0}.
\end{equation}

Thus we have just established that the unitary intertwining operator
(\ref{a018}) in our case is the Fourier integral operator \cite{FIO} and it acts on elements of the plane-wave basis of solutions of the free particle Dirac equation (\ref{a3}) in the following way:
\begin{equation}\label{Wf3}
\displaystyle
\tilde{{\varPsi}}_{q\zeta_{0}}(x)=U(x,x_{0}){\varPsi}_{{0}}(x) =
\frac{1}{2}\sum\limits_{\zeta = \pm 1} e^{-i((P_{\zeta}-q)
x)}\sqrt{|J_{\zeta}(q)|}{( 1 - \zeta \gamma^5
\gamma_{\mu}{S}_{{\mathrm{tp}}}^{\mu}(q))}{\varPsi}_{{0}}(x).
\end{equation}
\noindent Action of the inverse operator is defined by the formula
\begin{equation}\label{Wf4}
\displaystyle {\varPsi}_{{0}}(x)=
U^{-1}(x,x_{0})\tilde{{\varPsi}}_{q\zeta_{0}}(x) =
\frac{1}{2}\sum\limits_{\zeta = \pm 1} e^{i((P_{\zeta}-q)
x)}\frac{1}{\sqrt{|J_{\zeta}(q)|}}{( 1 - \zeta \gamma^5
\gamma_{\mu}{S}_{{\mathrm{tp}}}^{\mu}(q))}\tilde{{\varPsi}}_{q\zeta_{0}}(x).
\end{equation}
\noindent Since the intertwining operator is defined on the elements of the basis,
its action on an arbitrary solution is defined as well. Hence, the explicit
form of this operator as a function of coordinates and
differential operators can be easily obtained.

Note that an attempt to construct intertwining operator for equation
(\ref{6}) was undertaken in \cite{RL} for the case where only
parameter $f^{\mu}$ is nontrivial. The result of the action of the operator suggested in \cite{RL} on plane-wave solutions of the Dirac equation for a free
particle coincides with solutions obtained in our
previous work \cite{SpinLight} for this particular case, and which
can be derived from (\ref{Wf1}), if one sets $F^{\mu\nu}=0$. We should emphasize that solutions (\ref{Wf1}) do not form an orthogonal basis. As a
consequence of this fact, the intertwining operator suggested in
\cite{RL} is not unitary with regard to the standard scalar product
(\ref{a010}).

The case of a massless neutrino is quite special. Equation (\ref{103})
obviously does not hold
in the limit $m \rightarrow 0$
because massless particle helicity is conserved, while the BMT equation
describes spin rotation.
To find wave functions of massless neutrino, one  must take into account that from the mathematical point of view, the small
group of representation of the
Poincar\'{e} group for a massive particle differ from that for a
massless particle  (see, for example, \cite{BoLoTO}).
The small group for a
massive particle is the rotation group of three-dimensional Euclidian
space $O_{+}(3)$, and nontrivial magnetic and electric
moments are allowed. In contrast to that, the small
group for a massless particle is the movement group of Euclidian plane
$E(2)$, which outlaws nonzero magnetic moment as
well as electric one or other nontrivial coefficients $\mu_{n}$ and
${\varepsilon '}_{\!n}$ in equation (\ref{1.0}).  Therefore, one has to put
$\mu_{0}=0$ in equation (\ref{6}).
Then instead of (\ref{Wf}), we obtain
\begin{equation}\label{Wfx}
    \varPsi_{q\zeta}(x)|_{m=0} = e^{- i(qx) (1+\zeta \eta)}e^{-i(fx)(1-\zeta)/2}|1+\zeta \eta|
    (1 - \zeta\gamma^{5})\gamma^{\mu}{q}_{\mu}\psi_0,
\end{equation}
\noindent for the neutrino, and
\begin{equation}\label{Wfxx}
    \varPsi_{q\zeta}(x)|_{m=0} = e^{i(qx) (1+\zeta \eta)}e^{-i(fx)(1+\zeta)/2}|1+\zeta \eta|
    (1 + \zeta\gamma^{5})\gamma^{\mu}{q}_{\mu}\psi_0,
\end{equation}
\noindent for the antineutrino. In these equations $\eta=\frac{f^{2}}{2(fq)}$ and $q^{2}=0$.

\section{Operators of observables}\label{IV}

Let us find the explicit forms of the
kinetic momentum operators ${\mathfrak{p}}^{\mu}$ and the spin projection operator ${\mathfrak{S}}_{{\mathrm{tp}}}$ in the coordinate representation. For this purpose we might exploit formulas (\ref{a21});
however, we follow a simpler way.

The obtained solutions (\ref{Wf}) are classified by eigenvalues of the operators
${\mathfrak{p}}^{\mu}$ and ${\mathfrak{S}}_{{\mathrm{tp}}}$, so
\begin{equation}\label{operact} {\mathfrak{p}}^\mu
\varPsi_{q\zeta}(x)=q^\mu \varPsi_{q\zeta}(x), \quad
{\mathfrak{S}}_{{\mathrm{tp}}} \varPsi_{q\zeta}(x)=\zeta \varPsi_{q\zeta}(x).
\end{equation}
\noindent Since solutions (\ref{Wf}) are also eigenfunctions of the canonical
momentum operator $p^{\mu}=i\partial^{\mu}$ with eigenvalues
$P^{\mu}_{\zeta}$, we have
\begin{equation}\label{operP}
    {{p}}^{\mu}\varPsi_{q\zeta}(x)={{P}}^{\mu}_{\zeta}\varPsi_{q\zeta}(x).
\end{equation}
\noindent Now we should express eigenvalues of the kinetic momentum
operator ${q^{\mu}}$ in terms of eigenvalues of the canonical
momentum operator ${P}^{\mu}_{\zeta}$. From (\ref{P2}) we have
\begin{equation}\label{z4}
\begin{array}{c}
\displaystyle  q^\mu =\Bigg[{\tilde{P}}^{\mu}+ f^{\mu} \frac{\zeta
\sqrt{(\varphi q)^2-m^2\varphi^2}}{2(\varphi q)}
    +\varphi^\mu \,\frac{\zeta (f \varphi) m^2}
    {2(\varphi q)\sqrt{(\varphi
    q)^2-m^2\varphi^2}}\Bigg]\\[12pt]
\times\displaystyle \Bigg[1+\zeta\frac{(f\varphi)}
    {2\sqrt{(\varphi q)^2-m^2\varphi^2}}
    \Bigg]^{-1}.
\end{array}
\end{equation}
\noindent With the help of relations from Appendix \ref{ur} this can be
rewritten as
\begin{equation}\label{z13.x}
\begin{array}{l}
q^{\mu}=\displaystyle {\tilde{P}}^{\mu}+\frac{
    {\tilde{P}}^{\mu}(\tilde{\Phi}f)
    - f^{\mu}
(f\tilde{P})/2
    -2mH^{\mu\nu}\tilde{\Phi}_{\nu}
    }
{\tilde{P}^2 - m^2 + f^{2}/4+2\mu_{0}^{2}
    I_1-(\tilde{\Phi}f)}.
\end{array}
\end{equation}
\noindent The vector of total polarization in terms of the new variable
is
\begin{equation}\label{z15}
\begin{array}{c}
\displaystyle {S}_{{\mathrm{tp}}}^\mu(q) = \Delta\frac{q^\mu
(\tilde{\Phi}\tilde{P})/m -\tilde{\Phi}^{\mu} m }{\sqrt
{(\tilde{\Phi}\tilde{P})^2-m^2\tilde{\Phi}^2}}.
\end{array}
\end{equation}
Here $\tilde{P}, \tilde{\Phi}$ and $\Delta$ are given by
(\ref{bbb}).

Because of (\ref{operP}) we can interpret ${{P}}^{\mu}_{\zeta}$  as a
result of action of operator ${p}^\mu=i\partial^{\mu}$ on the wave
function. So by changing ${\tilde{P}}^\mu\Rightarrow {{p}}^\mu-f^\mu/2$
and $\tilde{\Phi}^\mu\Rightarrow
f^\mu/2+H^{\mu\nu}(p_\nu-f_\nu/2)/{m}$ in formulas (\ref{z13.x}),
(\ref{z15}), we obtain  kinetic momentum operator ${\mathfrak
p}^\mu$ and spin projection operator ${\mathfrak
S}_{{\mathrm{tp}}}=-\gamma^{5}\gamma_{\mu}S^{\mu}_{{\mathrm{tp}}}(q)$ in the explicit form. These operators are pseudodifferential ones \cite{FIO} and are
determined on the solutions of equation (\ref{6}) with fixed mass
$m$.

To extend the domain of the definition of constructed operators, we need
to replace mass $m$ in (\ref{z13.x}) and (\ref{z15}) by the matrix
operator from equation (\ref{6}):
\begin{equation}\label{z14}
\begin{array}{c}
\displaystyle m= \gamma^{\mu}{{{p}}}_{\mu}  -
\frac{1}{2}\gamma^{\mu}{f}_{\mu}(1 + \gamma ^5) - \frac{i}{2}\mu_{0}F^{\mu
\nu }\sigma _{\mu \nu },
\\[8pt]\displaystyle m^{2}=
\displaystyle {(p-f/2)}^{2}-f^{2}/4 +2\mu_{0}^{2}I_{1}
+\gamma^{5}\sigma^{\mu\nu}{f}_{\mu}{p}_{\nu}
+H^{\mu\nu}\gamma_{\mu}f_{\nu}(1+\gamma^{5})+
2\gamma^{5}H^{\mu\nu}\gamma_{\mu}{p}_{\nu}.
\end{array}
\end{equation}
\noindent In this way we obtain the covariant form for ${\mathfrak
p}^\mu$ and ${\mathfrak S}_{{\mathrm{tp}}}$.

Unfortunately, the result of this substitution cannot be written
as a compact formula, so we do not present it here. However, even
if we do not know a covariant form of the operator ${\mathfrak p}^\mu$,
we may conclude that on the solutions of equation (\ref{6}) the relations,
\begin{equation}\label{z1}
{\mathfrak p}^{2}=m^{2},\quad \gamma^\mu{\mathfrak p}_\mu=m,
\end{equation}
\noindent should hold. The first equality in (\ref{z1}) is obvious; to proof the
second one, recall that according to (\ref{a21})
\begin{equation}\label{z1.0}
m=U(x,x_{0})\gamma^{\mu}p_{\mu}U^{-1}(x,x_{0})=
  U(x,x_{0})\gamma^{\mu}U^{-1}(x,x_{0}){\mathfrak{p}}_{\mu}=
  (\gamma^{\mu}+\varGamma^{\mu}(q))\,{\mathfrak{p}}_{\mu}.
\end{equation}
\noindent From formulas (\ref{Wf3}) and (\ref{Wf4}) it follows that
$\varGamma^{\mu}(q) \sim S_{{\mathrm{tp}}}^{\mu}(q)$, and as a consequence
$\varGamma^{\mu}(q){\mathfrak{p}}_{\mu}=0 $. So equation (\ref{6})
may be represented as
\begin{equation}\label{z3}
(\gamma^{\mu}{{\mathfrak{p}}_{\mu}} -m)\varPsi(x) = 0.
\end{equation}

Consider now special cases where the presented technique looks quite
clear. Discuss the influence on the
neutrino dynamics of the electromagnetic field alone, i.e. assume that  $f^{\mu}=0$. In this case
eigenvalues of the kinetic momentum operator ${q^{\mu}}$ would be
expressible in terms of eigenvalues of the canonical momentum
operator ${P}^{\mu}_{\zeta}$ in the following way:
\begin{equation}\label{z24.0}
q^{\mu}
=P^{\mu}_{\zeta}-\frac{2H^{\mu\alpha}H_{\alpha\nu}P^{\nu}_{\zeta}}
{P^{2}_{\zeta}- m^{2}+2\mu_{0}^{2}I_{1}}.
\end{equation}
\noindent Then the covariant form of the kinetic momentum operator is
\begin{equation}\label{z25}
{\mathfrak{p}}^{\mu}=p^{\mu}+
\frac{H^{\mu\alpha}H_{\alpha\nu}p^{\nu}}
{\sqrt{p^{\beta}H_{\beta\alpha}H^{\alpha\rho}p_{\rho}}}\,\tilde{{\mathfrak{S}}}_{{\mathrm{tp}}},
\end{equation}
\noindent and the spin projection operator ${\mathfrak{S}}_{{\mathrm{tp}}}$ is
defined by the formula
\begin{equation}\label{z22}
{\mathfrak{S}}_{{\mathrm{tp}}} ={\mathrm{sign}}
\left(1+\frac{2\mu_{0}^{2}I_{1}}
{\sqrt{p^{\beta}H_{\beta\alpha}H^{\alpha\rho}p_{\rho}}}
\,\tilde{{\mathfrak{S}}}_{{\mathrm{tp}}}\!\right)
\tilde{{\mathfrak{S}}}_{{\mathrm{tp}}}.
\end{equation}
\noindent Here
\begin{equation}\label{z23}
\tilde{{\mathfrak{S}}}_{{\mathrm{tp}}}=\frac{\gamma^{5}\gamma_{\mu}H^{\mu\nu}p_{\nu}}
{\sqrt{p^{\beta}H_{\beta\alpha}H^{\alpha\rho}p_{\rho}}}.
\end{equation}
\noindent Thus operator ${\mathfrak{S}}_{{\mathrm{tp}}} $ has a simple physical
meaning. It characterizes a particle spin projection on the
direction of the magnetic field in the rest frame of the particle.

Vice versa, when the electromagnetic field is absent, but $f^\mu$ is
nontrivial, we have
\begin{equation}\label{z13.xx}
\begin{array}{l}
q^{\mu}=\displaystyle {P}_{\zeta}^{\mu}-\frac{1}{2}f^{\mu}+\frac{
    {P}_{\zeta}^{\mu}f^{2}
    - f^{\mu}
(f{P}_{\zeta})} {2({P}_{\zeta}^2 -({P}_{\zeta}f)- m^2)},
\end{array}
\end{equation}
\noindent and the covariant forms of the kinetic momentum and spin
projection operators are
\begin{equation}\label{z18.0}
{\mathfrak{p}}^{\mu}=p^{\mu}-
\frac{f^{\mu}}{2}-\gamma^{5}\frac{p^{\mu}f^{2}-f^{\mu}(fp)
}{2((pf)^{2}-p^{2}f^{2})}\sigma^{\alpha\nu}{f}_{\alpha}{p}_{\nu},
\end{equation}
\begin{equation}\label{z23.0}
{\mathfrak{S}}_{{\mathrm{tp}}}=\frac{\gamma^{5}\sigma^{\mu\nu}{f}_{\mu}{p}_{\nu}
}{\sqrt{(pf)^{2}-p^{2}f^{2}}}.
\end{equation}
\noindent Note that if the matter is at rest and nonpolarized
 $({\bf f}=0)$, then
\begin{equation}\label{z23.00}
{\mathfrak{S}}_{{\mathrm{tp}}}={\mathrm{sign}}(f^{0})\frac{({\bb \Sigma}\cdot{\bf p})}
{|{\bf p}|},
\end{equation}
\noindent in other words ${\mathfrak{S}}_{{\mathrm{tp}}}$ is equal to the
standard helicity operator up to the sign.

We can find now spin projection operators for nonstationary wave
functions (\ref{Wf1}) and (\ref{Wf2}). For this purpose, we introduce operators ${\mathfrak{S}}_{\pm}$ that act on the elements of system (\ref{Wf}) as follows:
\begin{equation}\label{sa}
{\mathfrak{S}}_{+} {\varPsi}_{q\zeta}(x) = \frac{(1-\zeta)}{2}{\varPsi}_{q(-\zeta)}(x),\quad {\mathfrak{S}}_{-} {\varPsi}_{q\zeta}(x) = \frac{(1+\zeta)}{2}{\varPsi}_{q(-\zeta)}(x).
\end{equation}
\noindent Then operators ${\mathfrak{S}}_{1}=\frac{1}{2}({\mathfrak{S}}_{+}+{\mathfrak{S}}_{-})$,
${\mathfrak{S}}_{2}=\frac{1}{2i}({\mathfrak{S}}_{+}-{\mathfrak{S}}_{-})$ and
${\mathfrak{S}}_{3}=\frac{1}{2}{\mathfrak{S}}_{{\mathrm{tp}}}$ correspond to elements of the Lie algebra of the $SU\!(2)$ group. Commutation relations for these operators are
\begin{equation}\label{W}
    [{\mathfrak{S}}_{i},{\mathfrak{S}}_{j}]=ie_{ijk}{\mathfrak{S}}_{k}.
\end{equation}
\noindent To determine the explicit realization of operators
${\mathfrak{S}}_{\pm}$ on eigenfunctions of operator ${\mathfrak{p}}^{\mu}$ let us choose the basis $S_{i}^{\mu}(q)$ (see (\ref{a110})) in the form
${S}^{\mu}_{{\mathrm{tp}}}(q), {S}^{\mu}_{1}(q), S^{\mu}_{2}(q)$.
Here ${S}^{\mu}_{{\mathrm{tp}}}(q)$ is defined by relation (\ref{P3}), the
other spacelike unit vectors are
\begin{equation}\label{z18}
{S}^{\mu}_{1
}(q)\!=\!\frac{S_{0}^{\mu}(q)+S_{{\mathrm{tp}}}^{\mu}(q)(S_{0}(q)
S_{{\mathrm{tp}}}(q))}
{\sqrt{1-(S_{0}(q)S_{{\mathrm{tp}}}(q))^{2}}},\quad {S}^{\mu}_{2
}(q)\!=\!\frac{e^{\mu\nu\rho\lambda}q_{\nu}S_{0\rho}(q)S_{{{\mathrm{tp}}}\lambda}(q)}
{m\sqrt{1-(S_{0}(q)S_{{\mathrm{tp}}}(q))^{2}}}.
\end{equation}
\noindent As a result we have
\begin{equation}\label{zz18}
{\mathfrak{S}}_{\pm}=-\frac{1}{2}\frac{\sqrt{|J_{\zeta =\pm 1}(q)|}}{\sqrt{|J_{\zeta =\mp 1}(q)|}}e^{\pm 2i\theta}\gamma^{5}\gamma_{\mu}\left({S}^{\mu}_{1}(q) \pm i{S}^{\mu}_{2}(q)\right),
\end{equation}
\noindent where
\begin{equation}\label{g3}
\theta = (Nx)\sqrt {(\varphi q)^2 -
    \varphi^2 m^2}/m.
\end{equation}

Operators
${\mathfrak{S}}_{{\mathrm{tp}}}$ and ${\mathfrak{S}}_{\pm}$ are integrals of motion. So the spin projection operator $\tilde{\mathfrak{S}}_{0}$ that has eigenfunctions (\ref{Wf2}) and eigenvalues $\zeta_{0}= \pm 1$ is a linear combination of these operators.
We can rewrite wave functions (\ref{Wf2}) in the form
\begin{equation}\label{Wf20}
\displaystyle \tilde{\varPsi}_{q\zeta_{0}}(x) =\sqrt{1-\zeta_{0}
(S_{0}(q)S_{{\mathrm{tp}}}(q))}\;\frac{{\varPsi}_{q\zeta=1}(x)}{\sqrt{2}}+
\zeta_{0}\sqrt{{1+\zeta_{0}(S_{0}(q)S_{{\mathrm{tp}}}(q))}}
\;\frac{{\varPsi}_{q\zeta=-1}(x)}{\sqrt{2}}.
\end{equation}
\noindent So it is obvious that
\begin{equation}\label{z19.1}
\begin{array}{c}
\displaystyle{\tilde{\mathfrak{S}}}_{0}=-(S_{0}(q)S_{{\mathrm{tp}}}(q))
{\mathfrak{S}}_{{\mathrm{tp}}}
+\displaystyle \sqrt{1-(S_{0}(q)S_{{\mathrm{tp}}}(q))^{2}}\,\left[\,{\mathfrak{S}}_{+}+
{\mathfrak{S}}_{-}\right].
\end{array}
\end{equation}

Similarly to (\ref{z19.1}), one can construct the integral of motion ${\mathfrak{S}}_{0}$
with eigenfunctions (\ref{Wf1}) and eigenvalues $\zeta_{0}= \pm 1$. Since
\begin{equation}\label{Wf10}
\displaystyle {\varPsi}_{q\zeta_{0}}(x) =\sqrt{1-\zeta_{0}
(S_{0}(q)S_{{\mathrm{tp}}}(q))}\;\frac{{\varPsi}_{q\zeta=1}(x)}
{\sqrt{2|J_{\zeta =+ 1}(q)|}}+
\zeta_{0}\sqrt{{1+\zeta_{0}(S_{0}(q)S_{{\mathrm{tp}}}(q))}}
\;\frac{{\varPsi}_{q\zeta=-1}(x)}{\sqrt{2|J_{\zeta =- 1}(q)|}},
\end{equation}
\noindent we have
\begin{equation}\label{z19}
\begin{array}{c}
\displaystyle{\mathfrak{S}}_{0}=-(S_{0}(q)S_{{\mathrm{tp}}}(q))
{\mathfrak{S}}_{{\mathrm{tp}}}
+\displaystyle \sqrt{1-(S_{0}(q)S_{{\mathrm{tp}}}(q))^{2}}\left[\frac{\sqrt{|J_{\zeta =- 1}(q)|}}{\sqrt{|J_{\zeta =+ 1}(q)|}}\,{\mathfrak{S}}_{+}+
\frac{\sqrt{|J_{\zeta =+ 1}(q)|}}{\sqrt{|J_{\zeta =- 1}(q)|}}\,{\mathfrak{S}}_{-}\right].
\end{array}
\end{equation}
\noindent Or, in the other form,
\begin{equation}\label{z19.2}
\begin{array}{c}
\displaystyle{\mathfrak{S}}_{0}=
\gamma^{5}\gamma_{\mu}\left\{(S_{0}(q)S_{\mathrm{tp}}(q)){{S}}_{\mathrm{tp}}^{\mu}(q)
-\displaystyle \sqrt{1-(S_{0}(q)S_{\mathrm{tp}}(q))^{2}}
\left[\cos2\theta\,{S}^{\mu}_{1}(q)
-\sin2\theta\,{S}^{\mu}_{2}(q)\right]\right\}.
\end{array}
\end{equation}
\noindent Note that operator (\ref{z19}) is not a self-adjoint operator with respect
to the standard scalar product (\ref{a010}). It seems quite natural,
since wave functions (\ref{Wf1}) do not form an orthogonal system. However, the
system of wave functions is orthonormalized to the condition ``one particle in the unit volume.'' In this sense wave functions (\ref{Wf1}) minimize the uncertainty relation which is due to (\ref{W})
\begin{equation}\label{uuu}
    \langle({\mathfrak{S}}_{1}-
    \langle{\mathfrak{S}}_{1}\rangle)^{2}\rangle
    \langle({\mathfrak{S}}_{2}-
    \langle{\mathfrak{S}}_{2}\rangle)^{2}\rangle=\frac{1}{4}
    \langle{\mathfrak{S}}_{3}\rangle^{2}.
\end{equation}
\noindent Therefore, these wave functions describe spin-coherent states of the neutrino
(on properties of coherent states see, for example, \cite{MM}). The given system of spin-coherent states is parametrized by four-vector $S^{\mu}_{0}(q)$.

\section{Quasiclassical interpretation}\label{VV}

Let us discuss in more detail the physical meaning of spin-coherent
states of the neutrino. For this purpose we construct vector and axial
currents with  the help of (\ref{Wf1}). The vector current is
\begin{equation}\label{g1}
j^{\mu}_{V}=\bar{\varPsi}(x)\gamma^{\mu}{\varPsi}(x)=q^{\mu}/q^{0},
\end{equation}
\noindent and we see that (\ref{Wf1}) describes neutrino propagation
with the constant velocity ${\bf{v}}_{{\mathrm{gr}}} ={\bf{q}}/q^{0}$. The axial
current is
\begin{equation}\label{ggg1}
j^{\mu}_{A}=\bar{\varPsi}(x)\gamma^{5}\gamma^{\mu}
{\varPsi}(x)=\zeta_{0}\frac{m}{q^{0}}S^{\mu}.
\end{equation}
\noindent Here
\begin{equation}\label{g2}
S^{\mu}=-S_{{\mathrm{tp}}}^{\mu}(q)(S_{0}(q)S_{{\mathrm{tp}}}(q))
+\left[S_{0}^{\mu}(q)+S_{{\mathrm{tp}}}^{\mu}(q)
(S_{0}(q)S_{{\mathrm{tp}}}(q))
\right]\cos2\theta
-\frac{1}{m}e^{\mu\nu\rho\lambda}q_{\nu}S_{0\rho}(q)
S_{{{\mathrm{tp}}}\lambda}(q)\sin2\theta,
\end{equation}
\noindent where $\theta$ is determined by (\ref{g3}). As expected
(see (\ref{V}) and (\ref{A})), vector and axial currents coincide
with the solutions of the BMT equation, if the proper time is defined as
in (\ref{xx}).

The three-vector of spin $\bb{\zeta}$ can be expressed in terms of the
four-vector $S^{\mu}$ components as
\begin{equation}\label{ad201}
   \bb{\zeta} ={\bf{S}}-\frac{{\bf{q}}S^0}{q^{0}+m}.
\end{equation}
Then for $\bb{\zeta}$ we have
\begin{equation}\label{ad201x}
   \bb{\zeta} ={\bb \zeta}_{{\mathrm{tp}}}({\bb \zeta}_{0}\cdot{\bb \zeta}_{{\mathrm{tp}}})+
   [{\bb \zeta}_{0}- {\bb \zeta}_{{\mathrm{tp}}}({\bb \zeta}_{0}\cdot{\bb \zeta}_{{\mathrm{tp}}})]
   \cos2\theta
   -[{\bb \zeta}_{{\mathrm{tp}}}\times
   {\bb \zeta}_{0}]\sin2\theta.
\end{equation}
\noindent Expression (\ref{ad201x})
has a simple quasiclassical
interpretation.

The antisymmetric tensor $G^{\mu \nu}$ (see equation (\ref{5})) can
be written in the standard form
\begin{equation}
G^{\mu \nu}= (\mu_{0}{\bf P}, \mu_{0}{\bf M}\,), \label{g6}
\end{equation}
where
\begin{equation}\label{g7}
{\bf M}= (f^{0} {\bf q}-q^{0}{\bf f})/(2\mu_{0}m), \quad {\bf P}=-
\big[{\bf q \times {\bf f}}\big]/(2\mu_{0}m).
\end{equation}
\noindent Vectors ${\bf P}$ and ${\bf M}$ are analogous to the
polarization and the magnetization vectors of medium. Note that the
substitution $F^{\mu\nu}\Rightarrow F^{\mu\nu}+G^{\mu \nu}/\mu_{0}$ implies
that the magnetic ${\bf H}$ and electric ${\bf D}$ fields are
shifted by the vectors ${\bf M}$ and ${\bf P}$, respectively (we use
here notation common in electrodynamics of continuous media
\cite{LL})
\begin{equation}\label{g8}
{\bf H} \Rightarrow {\bf B} ={\bf H} +{\bf M}, \quad {\bf D}
\Rightarrow {\bf E}= {\bf D} -{\bf P}.
\end{equation}
\noindent  Thus restriction (\ref{03}) in the explicit form is
\begin{equation}\label{fields}
({\bf{E}}\cdot{\bf{f}})=0, \quad
  {\bf{E}}f^0-[{\bf{B}}\times{\bf{f}}]=0.
\end{equation}
\noindent This means that the Lorentz force and moment of force
acting on matter are equal to zero, i.e. matter is at equilibrium
state within the accuracy of our consideration. In particular,
the vector of polarization of matter is parallel to the magnetic field if
the matter is at rest.

In the rest frame of the particle, equation (\ref{x4}) can
be written in the form
\begin{equation}\label{g11}
\dot{{\bb \zeta}}={2}\mu_{0} \big[ {{\bb\zeta} \times {\bf B}_0} \big],
\end{equation}
\noindent where the spin vector $\bb\zeta$ is related to four-vector
$S^{\mu}$ by equation (\ref{ad201}) and the value ${\bf B}_0$  is
the effective magnetic field in the neutrino rest frame.  This field
can be expressed in terms of quantities determined in the laboratory
frame,
\begin{equation}\label{g12}
\begin{array}{l}
{\bf B}_0= \displaystyle\frac{1}{m}\left[q^{0}{\bf B} -\big[{\bf
q}\times{\bf E}\big]-\frac{{\bf q}({\bf q}\cdot{\bf
B})}{q^{0}+m}\right]\\[12pt]\displaystyle =\frac{1}{m}\left[q^{0}{\bf H}
-\big[{\bf q}\times{\bf D}\big]-\frac{{\bf q}({\bf q}\cdot{\bf
H})}{q^{0}+m}+\frac{\bf q}{2\mu_{0}} \Big(f^{0}-\frac{({\bf q}\cdot{\bf
f})}{q^{0}+m}\Big)\right] -\frac{\bf f}{2\mu_{0}}.
\end{array}
\end{equation}
\noindent We see that the neutrino spin precesses around the direction ${\bf
B}_{0}$ with the frequency $\omega = 2m|{\bf B}_{0}|/q^{0}$,
the angle between ${\bf B}_{0}$ and the vector of spin being
$\vartheta = \arccos({\bb{\zeta}}_{0}\cdot{\bb{\zeta}}_{{\mathrm{tp}}})$. The spin vector direction corresponding to stationary
states $\bb{\zeta}_{{\mathrm{tp}}}$ is connected to the effective magnetic
field as follows:
\begin{equation}\label{g13}
{\bb{\zeta}}_{{\mathrm{tp}}}=\frac{{\bf B}_0}{|{\bf B}_0|}.
\end{equation}
\noindent This fact explains in a simple way the stationarity of states with
$S^{\mu}_{0}(q)=S^{\mu}_{{\mathrm{tp}}}(q)$.

Let us introduce a flight length $L$ and a spin oscillation length
$L_{{\mathrm{osc}}}$ of the particle, remember that these are related by
$\theta =\pi L/L_{{\mathrm{osc}}}$ and that the scalar product $(Nx)= \tau$
may be interpreted as the proper time of a particle. The spin oscillation
length is
\begin{equation}\label{g4}
L_{{\mathrm{osc}}} = \frac{2\pi{|\bf{q}|}}{\sqrt {(fq)^2 - f^{2}
m^2 - 4m
H^{\mu\nu}f_{\mu}q_{\nu}+4H^{\mu\alpha}
H_{\alpha\nu}q_{\mu}q^{\nu}}}.
\end{equation}

We can now write the probability for the neutrino which was arisen
with polarization ${\bb \zeta}_{0}$ to change the polarization to
${-\bb \zeta}_{0}$ after traveling some distance $L$. This is
\begin{equation}\label{g14}
    W= \sin^{2}\vartheta
    \sin^{2}(\pi L/L_{{osc}}).
\end{equation}
\noindent It is clear that when $({\bb{\zeta}}_{0}\cdot{\bb
{\zeta}}_{{\mathrm{tp}}})=0$ or $\vartheta =\pi/2$, the probability can be
equal to unity and the resonance takes place.

If the neutrino has the fixed helicity in the initial state
\begin{equation}\label{g003}
S_{0}^{\mu}(q) =\frac{1}{m}\left\{{|\bf q|},
q^{0}{\bf q}/{|\bf q|}\right\},\quad
{\bb{\zeta}}_{0}= \frac{{\bf q}}{|{\bf q}|},
\end{equation}
\noindent then $\sin^{2}\vartheta = 1-({\bf{B}}_{0}\cdot{\bf
q})^{2}/|{\bf{B}}_{0}|^{2}|{\bf q}|^{2}$ and formula (\ref{g14})
simplifies to the result widely discussed in the literature.
This fact is not surprising. As was mentioned in Introduction, the Schr\"{o}dinger-type equation with an effective Hamiltonian,
which was used in papers \cite{theor} for obtaining this
result, is merely the BMT equation in the spinor representation.

\section{Discussion and conclusions}\label{VI}

Let us compare our results with those of the standard
quantum-mechanical approach to this problem. Since the neutrino
behavior in dense matter under the influence of magnetic field, to
the best of our knowledge, was not investigated before, we consider
the neutrino propagation in a constant homogeneous magnetic field
alone.

In the studies of the influence of a stationary pure magnetic field on the
neutrino spin rotation in the pioneer paper \cite{Fu}, as well as in
other papers, the stationary solutions
$\varPsi_{p{\zeta}}(x)$ first found in \cite{1} were used as
the wave functions of a particle. These solutions are the
eigenfunctions of the canonical momentum operator $p^{\mu}$ and of
the spin projection operator ${{\mathfrak{S}}}_{{\mathrm{tp}}}$ (see (\ref{z22})).
The description of the neutrino spin rotation there is based on
solving the Cauchy problem where the initial condition is chosen in
such a way that the mean value of neutrino helicity is equal to $\pm
1$. It was taken for granted that the solution of the Cauchy problem can be
expressed as a linear combination of the above-mentioned wave
functions:
\begin{equation}\label{zx}
\varPsi(x)=\sum_{{\zeta} = \pm
1}c_{{\zeta}}(p)\varPsi_{p{\zeta}}(x).
\end{equation}
\noindent However, such an assumption is incorrect.
The point is that, once in a pure state the mean value of some
spin operator is equal to  $\pm 1$, then this state is described by
an eigenfunction of this operator. In the general case, the construction of the eigenfunction of the spin projection
operator as a superposition of only positive-energy
solutions of  equation (\ref{6}) is possible only when
this spin projection operator commutes with the operator of the sign of the energy.
The standard helicity operator $({\bb \Sigma}\cdot{\bf p})/|{\bf p}|$ does
not feature it.

The given phenomenon is a sort of the famous Klein paradox
\cite{Klein}. To avoid the indicated difficulties, in relativistic
quantum mechanics only self-adjoint
operators in the subspace of wave functions with a fixed energy sign can
be treated as operators of observables.
The choice of integrals of motion as operators of observables is the
necessary condition to satisfy this requirement \cite{LanP}.

In the case considered the canonical momentum operator is an integral of motion. However, the conserved operator of the spin projection which should set
initial conditions to the Cauchy problem is uniquely --- up to the sign --- determined by the form of the Dirac--Pauli equation. This operator is ${{\mathfrak{S}}}_{{\mathrm{tp}}}$.
Therefore, it is impossible to
construct a wave function describing a neutrino with rotating spin
in the form of an eigenfunction of the canonical momentum operator
for its arbitrary eigenvalues.
The solutions similar to (\ref{zx}) can exist
when the special values of the canonical
momentum are chosen. So, if a neutrino moves parallel or
perpendicular to a constant homogeneous magnetic field,
eigenfunctions of the helicity
operator are the superpositions of positive-energy solutions alone \cite{BTZ}.

To solve the problem we abandon
the view that eigenvalues of the canonical momentum operator always impose
a direction of the particle propagation.
We found a self-adjoint operator ${\mathfrak{p}}^{\mu}$ which
can be interpreted as kinetic momentum operator of the particle and
obtained the complete orthonormal system of the solutions of
equation (\ref{6}) with elements which are eigenfunctions of the given
operator. On the base of this system we constructed solutions
describing the neutrino with rotating spin. So our results enable one to
treat a possible effect of the neutrino polarization change as a
real precession of the particle spin.

Consequently, the problem of
neutrino spin rotation in dense matter and in strong electromagnetic
fields is solved in full agreement with the basic principles of
quantum mechanics. Using the wave functions of orthonormal
basis (\ref{Wf}) or spin-coherent wave functions (\ref{Wf1}),
it is possible to calculate probabilities of various processes with
the neutrino in the framework of the Furry picture. When choosing
one or another type of the basis, it is necessary to take into
account that, due to the time-energy uncertainty, stationary
states of the neutrino can be generated only when the linear size of
the area occupied by the electromagnetic field and the matter is
comparable in the order of magnitude with the formation length of the process
--- the spin oscillation length in our case.

\acknowledgments

The authors are grateful to V. G. Bagrov, A. V. Borisov, Ya. N. Istomin, O. S.
Pavlova, A. E. Shabad, and V.~Ch.~Zhukovsky  for helpful discussions.

\bigskip

The work was supported in part by grants of the President of the
Russian Federation for leading scientific schools (Grant No. SS ---
65255.2010.2 and Grant No. SS ---
4142.2010.2).

\appendix

\section{}\label{uuur}

Let us find the equation for the description of the neutrino
behavior in dense matter and in electromagnetic fields at low-energy limit.
When weak interaction with background fermions is considered to be coherent, the behavior of mass states of any one-half spin lepton should be described by the Dirac-type equation,
\begin{equation}\label{1}
\left(i\gamma^{\mu}{\partial}_{\mu} + {\mathfrak V}_{{\mathrm{em}}} +
{\mathfrak V}_{{\mathrm{matter}}} - m\right)\varPsi(x) = 0.
\end{equation}
\noindent In this equation the term ${\mathfrak V}_{{\mathrm{em}}}$ describes interaction of
the particle with the electromagnetic field and the term ${\mathfrak
V}_{{\mathrm{matter}}}$ is responsible for weak interaction with matter.

Following paper \cite{F}, the nature of the interaction terms is
determined by the restrictions that the equation be Lorentz covariant
and gauge invariant; that the terms are linear in the
electromagnetic fields and integral characteristics of matter, i.e.
currents and polarizations of background particles; that terms do
not vanish in the limit of vanishing momentum of the particle; that
the charge and current distribution associated with the particle be
sufficiently localized that its interaction with slowly varying
electromagnetic fields and characteristics of matter may be
expressed in terms of the electromagnetic and matter potentials and
arbitrary high derivatives of these potentials evaluated at the
position of the particle. These assumptions lead to the term
${\mathfrak V}_{{\mathrm{em}}}$ in the form
\begin{equation}\label{1.0}
{\mathfrak
V}_{{\mathrm{em}}}=-\sum\limits_{n=0}^{\infty}\left[\varepsilon_{n}\gamma^{\mu}
{{\square}}^{n}A_{\mu}+\frac{i}{2}\mu_{n}\sigma^{\mu\nu}
{{\square}}^{n}F_{\mu\nu}+{\mu '}_{\!n}\gamma^{5}
(\gamma^{\mu}{\square}-\gamma^{\nu}\partial_{\nu}\partial^{\mu})
{{\square}}^{n}A_{\mu}+\frac{1}{2}{\varepsilon
'}_{\!n}\gamma^{5}\sigma^{\mu\nu} {{\square}}^{n}F_{\mu\nu}\right].
\end{equation}
\noindent Here $A^{\mu}$ is the potential of the external electromagnetic
field, $F^{\mu\nu}=\partial^{\mu} A^{\nu}-\partial^{\nu} A^{\mu}$ is
the electromagnetic field tensor,
${\square}=\partial^{\mu}\partial_{\mu}$ is the d'Alembert operator.
The constants $\mu_{n},{\mu'_{n}},\varepsilon_{n},{\varepsilon'_{n}}$
characterize the interaction, $\varepsilon_{0}$ is charge of the
particle, $\mu_{0},{\varepsilon'}_{\!0}$ are, respectively,
anomalous magnetic and electric moments, and ${\mu'}_{\!0}$ is an
anapole moment. The expression for the ${\mathfrak V}_{{\mathrm{matter}}}$ can
be found if  we replace $A_{\mu}$ in (\ref{1.0}) with a linear combination of the currents (\ref{3}) and of the polarizations (\ref{4})
of background fermions $f$ with the proper choice of coupling constants.

We have the minimal nontrivial generalization of the Dirac--Pauli equation
neglecting terms with derivatives higher than the second in (\ref{1.0}).
In the expression for ${\mathfrak V}_{{\mathrm{matter}}}$ we must hold only leading terms due to the proportionality of ${\square}A^{\mu}$ to the sum of charged particle currents. Further restrictions for (\ref{1}) depend on the sort of lepton and on the model of interaction.
The neutrino is a neutral particle, thus $\varepsilon_{0}=0.$ In the
framework of the standard model where it is assumed that the theory
is $T$ invariant and the neutrino interacts with  leptons and quarks through left
currents, its anomalous electric moment goes to zero
(${\varepsilon'}_{0}=0$) and the term describing direct interaction
of the neutrino with the currents contains multiplier $(1+\gamma^{5})$.
As a result we come to equation (\ref{6})
with an effective four-potential $f^\mu$ which is determined by (\ref{2}) and (\ref{rho}).

\section{}\label{ur}

For an arbitrary antisymmetric tensor $A^{\mu\nu}$,
its dual tensor ${}^{\star\!\!}A^{\mu\nu}=- \frac{1}{2}e^{\mu \nu
\rho \lambda } A_{\rho \lambda}$, and for any four-vectors
$g^{\mu},h^{\mu}$, such as $(gh)\neq 0$  the following relation takes place \cite{Bacry}:
\begin{equation}\label{A2.16}
A^{\mu\nu}(gh)=
-\left[g^{\mu}A^{\nu\rho}h_{\rho}-A^{\mu\rho}h_{\rho}g^{\nu}\right]
+{}^{\star}\left[h^{\mu}
{}^{\star\!\!}A^{\nu\rho}g_{\rho}-{}^{\star\!\!}
A^{\mu\rho}g_{\rho}h^{\nu}\right].
\end{equation}
\noindent This leads to the formula
\begin{equation}\label{A2.016}
\begin{array}{l}
g_{\mu}{}^{\star\!\!}A^{\mu}_{{\phantom{\nu}}\rho}{}^{\star\!\!}
A^{\rho}_{{\phantom{\nu}}\nu}g^{\nu}h^{2}+\left(g_{\mu}{}^
{\star\!\!}A^{\mu}_{{\phantom{\nu}}\nu}h^{\nu}\right)^{2}
-h_{\mu}A^{\mu}_{{\phantom{\nu}}\rho}
A^{\rho}_{{\phantom{\nu}}\nu}h^{\nu}g^{2}-\left(g_{\mu}
A^{\mu}_{{\phantom{\nu}}\nu}h^{\nu}\right)^{2}=
\\[8pt]=g_{\mu}{}^{\star\!\!}A^{\mu}_{{\phantom{\nu}}\rho}{}^{\star\!\!}
A^{\rho}_{{\phantom{\nu}}\nu}h^{\nu}(gh)-g_{\mu}A^{\mu}_{{\phantom{\nu}}\rho}
A^{\rho}_{{\phantom{\nu}}\nu}h^{\nu}(gh)=2(gh)^{2}I_{1}.
\end{array}
\end{equation}
\noindent Here $I_{1}= \frac{1}{4}A^{\mu\nu}A_{\mu\nu}= -\frac{1}{4}{}^{\star\!\!}A^{\mu\nu}{}^{\star\!\!}A_{\mu\nu}$ is the first invariant of the tensor $A^{\mu\nu}$.

Let $A^{\mu\nu}={}^{\star\!}F^{\mu\nu}$, $g^{\mu}=\varphi^{\mu}$, and $h^{\mu}=q^{\mu}$. Since  ${}^{\star\!}F^{\mu\nu}\varphi_{\nu} = -F^{\mu\nu}\varphi_{\nu} = 0$ (see (\ref{dop11}), (\ref{dop12})), then from (\ref{A2.16}) and (\ref{A2.016}) we get
\begin{equation}\label{A2.1}
    2H^{\mu\nu} (\varphi q)=H^{\mu\rho}q_\rho f^{\nu}-f^\mu H^{\nu\rho}q_{\rho}=m(\varphi^{\mu} f^{\nu}-f^{\mu}\varphi^{\nu}),
\end{equation}
\noindent and
\begin{equation}\label{A2.3}
\begin{array}{c}
    m^2((f\varphi)^2-f^2 \varphi^2)=f^2 {\cal N} + (f_\mu H^{\mu\nu} q_\nu)^2
    =(fq)f_{\mu}H^{\mu\alpha}H_{\alpha\nu}q^{\nu}=2(fq)^2\mu_{0}^{2} I_1,\\[8pt]\displaystyle
{\cal N} - {\tilde{\cal N}}=2 m^2 \mu_{0}^{2}I_{1}.
\end{array}
\end{equation}

Let $A^{\mu\nu}={}^{\star\!}F^{\mu\nu}$, $g^{\mu}=\tilde{\Phi}_{\mu}$, and $h^{\mu}=\tilde{P}^{\mu}$. Since ${}^{\star\!}F^{\mu\nu}\tilde{\Phi}^{\mu} = -F^{\mu\nu}\tilde{\Phi}_{\nu} = 0$, then from (\ref{A2.16}), (\ref{A2.016}) we get
\begin{equation}\label{z8}
\displaystyle  2H^{\mu\nu}{(\tilde{\Phi}\tilde{P})}=H^{\mu\rho}\tilde{P}_\rho f^{\nu}-f^\mu H^{\nu\rho}\tilde{P}_{\rho}={m}({\tilde{\Phi}^\mu
     f^\nu-f^\mu \tilde{\Phi}^\nu}),
\end{equation}
\noindent and
\begin{equation}\label{z13.4}
m^{2}((f\tilde{\Phi})^{2}-f^{2}\tilde{\Phi}^2
    )=f^2 \tilde{P}_{\mu}H^{\mu\alpha}H_{\alpha\nu}\tilde{P}^{\nu} + (f_\mu H^{\mu\nu} \tilde{P}_{\nu})^2
    =(f\tilde{P})f_{\mu}H^{\mu\alpha}H_{\alpha\nu}\tilde{P}^{\nu}
    =2(f\tilde{P})^{2}\mu_{0}^{2}I_{1}.
\end{equation}

It is possible to establish by direct calculations using (\ref{A2.3}) that

\begin{equation}\label{vvv}
\displaystyle  \frac{(\tilde{P}\tilde{\Phi})}{(q\varphi)}= \frac{(f\tilde{\Phi})}{(f\varphi)}= \left({1+\zeta\frac{f_{\mu}H^{\mu\nu}q_{\nu}/2m-2\mu_{0}^{2}
    I_1}
    {\sqrt{(\varphi q)^2-m^2 \varphi^2}}}\right).
\end{equation}
\noindent It follows from (\ref{A2.1}) and (\ref{z8}) that
\begin{equation}\label{vv}
\frac{\varphi^\mu}{(\varphi q)} =
\frac{\tilde{\Phi}^{\mu}}{(\tilde{\Phi}\tilde{P})},
\end{equation}
\noindent and from (\ref{A2.3}), (\ref{z13.4}), and (\ref{vvv}) that
\begin{equation}\label{z5}
\begin{array}{c}
\displaystyle {(\varphi q)^2-m^2\varphi^2}=\frac{(\varphi q)^2}{(\tilde{\Phi}
\tilde{P})^2}\,[{(\tilde{\Phi}\tilde{P})^2-m^2\tilde{\Phi}^2}],
\\[8pt]
\displaystyle  \frac{(\varphi q)}{\sqrt{(\varphi q)^2-m^2\varphi^2}}=\Delta
 \frac{(\tilde{\Phi}\tilde{P})}{\sqrt{(\tilde{\Phi}\tilde{P})^2-m^2\tilde{\Phi}^2}}.
\end{array}
\end{equation}

\section{}\label{rr}

The Lorentz equation for the four-velocity $u^{\mu} $ and the BMT
equation for the spin vector $S^{\mu}$ are
\begin{equation}\label{BMT3}
\begin{array}{c}
    \dot{u}^\mu=\displaystyle\frac{e}{m}F^{\mu\nu} u_{\nu},\\
    \dot{S}^\mu=\displaystyle\frac{e}{m}F^{\mu\nu} S_{\nu}+2\mu_{0} \left( g^{\mu\lambda} -
    u^{\mu}u^{\lambda}\right) F_{\lambda\nu} S^{\nu}.
\end{array}
\end{equation}
\noindent Since $u^{2}=1$ and $(Su) =0$, we may rewrite these equations as
\cite{Zw}
\begin{equation}\label{b5}
\dot{u}^\mu=\Omega^{\mu\nu} u_{\nu},\quad
\dot{S}^\mu=\Omega^{\mu\nu} S_{\nu},
\end{equation}
\noindent where
\begin{equation}\label{b6}
\Omega^{\mu\nu}=\displaystyle\frac{e}{m}F^{\mu\nu}+
2\mu_{0}(g^{\mu\alpha}-u^{\mu}u^{\alpha})F_{\alpha\beta}
(g^{\beta\nu}-u^{\beta}u^{\nu})
\end{equation}
\noindent is an antisymmetric tensor.

From (\ref{b5}) it is obvious that the evolution operators for the
Lorentz and BMT equations are the same and fulfill the relation
\begin{equation}\label{b7}
\dot{R}^{\mu\nu}(\tau,\tau_{0})=\Omega^{\mu}_{{\phantom{\mu}}\alpha}{R}^{\alpha\nu}
(\tau,\tau_{0}).
\end{equation}
\noindent From (\ref{V}) and (\ref{A})  it  follows that
\begin{equation}\label{VA1}
j _{V}^{\mu}(\tau)={R}^{\mu}_{{\phantom{\mu}}\nu}(\tau,\tau_{0})j _{V}^{\nu}(\tau_{0})=\bar{\varPsi}(\tau)\gamma^{\mu}\varPsi(\tau)=
\bar{\varPsi}(\tau_{0})U(\tau_{0},\tau)\gamma^{\mu}U(\tau,\tau_{0})\varPsi(\tau_{0}),
\end{equation}
\noindent and
\begin{equation}\label{VA2}
j _{A}^{\mu}(\tau)={R}^{\mu}_{{\phantom{\mu}}\nu}(\tau,\tau_{0})j _{A}^{\nu}(\tau_{0})=\bar{\varPsi}(\tau)\gamma^{5}\gamma^{\mu}\varPsi(\tau)=
\bar{\varPsi}(\tau_{0})U(\tau_{0},\tau)\gamma^{5}\gamma^{\mu}
U(\tau,\tau_{0})\varPsi(\tau_{0}).
\end{equation}

Therefore, the equation for the evolution operator of the
quasiclassical spin wave function takes the form
\begin{equation}\label{b8}
\dot{U}(\tau,\tau_{0})=Z{U}(\tau,\tau_{0}),
\end{equation}
\noindent where $Z$ must obey the relations
\begin{equation}\label{b9}
[\gamma^{\mu},Z]={\Omega}^{\mu\nu}\gamma_{\nu},\quad  [\gamma^{5}\gamma^{\mu},Z]=
{\Omega}^{\mu\nu}\gamma^{5}\gamma_{\nu}.
\end{equation}
\noindent  It is obvious that
\begin{equation}\label{b10}
Z=\frac{1}{4}\Omega_{\mu\nu}\sigma^{\mu\nu}.
\end{equation}
\noindent  Using the relations
\begin{equation}\label{x6}
\frac{1}{2} F^{\mu\nu}\sigma_{\mu\nu}=i\gamma^5 {}^{\star\!\!}F^{\mu\nu}
u_{\nu} \gamma_{\mu}{u}^{\alpha}\gamma_{\alpha}+F^{\mu\nu} u_{\nu} \gamma_{\mu
}{u}^{\alpha}\gamma_{\alpha},
\end{equation}
\noindent  and
\begin{equation}\label{x7}
F^{\mu\nu}\sigma_{\mu\nu}=i\gamma^{5}{}^{\star\!}F^{\mu\nu}\sigma_{\mu\nu},
\end{equation}
\noindent  we find
\begin{equation}\label{x8}
Z=i\gamma^{5}\left(\frac{e}{4m}F^{\mu\nu}\gamma_{\mu}\gamma_{\nu}  + \mu_{0}F^{\mu\nu }u_{\nu} \gamma_{\mu} \gamma^{\alpha}{u}_{\alpha}\right).
\end{equation}

We must take into account that the electric charge of neutrino $e$ is equal to zero. Replacing $\mu_{0}{}^{\star\!}F^{\mu\nu}$ to $\mu_{0}{}^{\star\!}F^{\mu\nu}+ (f^{\mu}u^{\nu}-u^{\mu}f^{\nu})/2$ (see (\ref{x4}) and (\ref{5})), and introducing the notation $q^{\mu}=mu^{\mu}$
we obtain equation (\ref{103}).

\section{}\label{ru}

Let us prove the system (\ref{Wf}) is orthonormal, i.e.
\begin{equation}\label{ort01}
 {\mathbb{N}}^{2}_{+} = \int\! d{{\bf
x}}\,\varPsi^{\dag}_{q'\zeta'}(x)\varPsi_{q\zeta}(x)=(2\pi)^{3}
\delta^{3}({\bf q}-{\bf q}')\delta_{{\zeta}{\zeta'}}
\end{equation}
\noindent for the solutions with the same signs of energy and
\begin{equation}\label{ort001}
 {\mathbb{N}}^{2}_{-} = \int\! d{{\bf
x}}\,\varPsi^{\dag}_{q'\zeta'}(x)\varPsi_{q\zeta}(x)=0
\end{equation}
\noindent for the solutions with the different signs of energy.
After integration we get
\begin{equation}\label{ort011}
{\mathbb{N}}^{2}_{+} = (2\pi)^{3}\sqrt{J_{\zeta'}(\pm q')}\sqrt{J_{\zeta}(\pm q)}
\left(\psi^{\pm}_{\zeta'}(q')\right)^{\dag}\psi^{\pm}_{\zeta}(q)
\delta^{3}({\bf P}_{\zeta'}(\pm q')-{\bf P}_{\zeta}(\pm q)),\\
\end{equation}
\begin{equation}\label{ort111}
{\mathbb{N}}^{2}_{-} = (2\pi)^{3}\sqrt{J_{\zeta'}(\pm q')}\sqrt{J_{\zeta}(\mp q)}
\left(\psi^{\pm}_{\zeta'}(q')\right)^{\dag}\psi^{\mp}_{\zeta}(q)
\delta^{3}({\bf P}_{\zeta'}(\pm q')-{\bf P}_{\zeta}(\mp q)),
\end{equation}
\noindent where the conventional eigenspinors
$\psi^{\pm}_{\zeta}(q)$ obey the equations $(\gamma^{\mu}q_{\mu}\mp
m)\psi^{\pm}_{\zeta} (q)=0$; $J_{\zeta}(q)$ is the Jacobian for
transition between the variables $q^\mu$ and $P^\mu_{\zeta}(q)$.

In the standard representation for the gamma matrices, we obtain explicitly \cite{BLP}
\begin{eqnarray*}
\psi^{+}_{\zeta}({{q}})&=&\frac{1}{\sqrt{2q^{0}(q^{0}+m)}}
\left(\begin{array}{c}
(q^{0}+m)\,\omega^{+}_{\zeta}\\ ({\bb{\sigma}}\cdot{\bf {q}})\,\omega^{+}_{\zeta}
\end{array}\right),\nonumber\\
\psi^{-}_{\zeta}({{q}})&=&\frac{1}{\sqrt{2q^{0}(q^{0}+m)}}
\left(\begin{array}{c}
({\bb{\sigma}}\cdot{\bf {q}})\,\omega^{-}_{\zeta}\\ (q^{0}+m)\,\omega^{-}_{\zeta}
\end{array}\right),
\end{eqnarray*}

\noindent where $\sigma_{i}$ are the Pauli matrices and factor $1/\sqrt{2q^{0}(q^{0}+m)}$ has been included for normalization (\ref{aaa3}).
These spinors satisfy the orthogonality relations
\begin{equation}\label{orthonorm1}
\left({\psi^{\pm}_{\zeta}}({{q}})\right)^{\dag}\gamma^{0} \psi^{\mp}_{\zeta'}({{q}})=0,
\quad
\left({\psi^{\pm}_{\zeta}}(q)\right)^{\dag} \psi^\pm_{\zeta'}({{q}})=\delta_{\zeta\zeta'},
\end{equation}
if $\omega^{\pm}_{\zeta}$ are nonvanishing, but otherwise arbitrary
two-component spinors which are chosen such that
$({\omega^{\pm}_{\zeta'}})^{\dag}{\omega^{\pm}_{\zeta}}=\delta_{\zeta\zeta'}$.
It is convenient to use the remaining uncertainty in
$\omega^{\pm}_{\zeta}$ to require $\psi^{\pm}_{\zeta}({{q}})$
to be eigenstates of the spin projection operator
\begin{equation}\label{}
-\gamma^{5}\gamma_{\mu}S^{\mu}_{{\mathrm{tp}}}(q)\psi^{\pm}_{\zeta}({{q}})
=\zeta\psi^{\pm}_{\zeta}({{q}}).
\end{equation}
\noindent For this purpose $\omega_{\zeta}^{\pm}$ should be
eigenspinors of three-dimensional spin projection operator (see
(\ref{ad201}))
\begin{equation}\label{}
    ({\bb{\sigma}}\cdot{\bb{\zeta}}_{{\mathrm{tp}}})\,
    \omega_{\zeta}^{\pm}=\pm\zeta\omega_{\zeta}^{\pm}.
\end{equation}

To calculate spinors $\omega^{\pm}_{\zeta}$ we have an opportunity
to choose a special reference frame. Let us  take the reference
frame where $ f^\mu=\{ f,0,0,0 \} $, ${\bf{H}}=(0,0,H)$, and
${\bf{E}}=0$. It is possible, if $f^\mu$ is timelike four-vector
and the relation $F^{\mu\nu}f_\nu=0$ is fulfilled.

In this reference
frame the vector of the total polarization is
\begin{equation}
    S^\mu_{{\mathrm{tp}}}(q)=\frac{1}{mR_{\pm}}\Big\{ q_{\bot}+q_{\scriptscriptstyle{\parallel}}
    {\tilde{q}}_{\pm}/q_{\bot},\,
    q^0\cos\phi , \,  q^0\sin\phi , \,
    q^0{\tilde{q}}_{\pm}/q_{\bot}\Big\},
\end{equation}
\noindent and
\begin{equation}
    {\bb{\zeta}}_{{\mathrm{tp}}}=
    \frac{1}{q_{\bot}R_{\pm}} \left\{
    {\bf{q}}\left[ 1 \mp \displaystyle \frac{2\mu_{0}Hq_{\scriptscriptstyle{\parallel}}}{f(q^0+m)} \right]
    \pm 2\mu_{0}{\bf{H}}\,\frac{q^{0}}{f}
    \right\},
\end{equation}
\noindent where
\begin{equation}
R_{\pm}={\mathrm{sign}}(f)\sqrt{ (1+4\mu_{0}^{2}H^{2}/f^{2})+(\tilde{q}_{\pm}/q_{\bot})^{2}}.
\end{equation}
\noindent Here we use the notation
\begin{equation}
    \sqrt{({q^{1}})^{2}+({q^{2}})^{2}}=q_{\bot},\; q^1=q_{\bot}\cos\phi, \; q^2=q_{\bot}\sin\phi,\; q^{3}= q_{\scriptscriptstyle{\parallel}},\;  q_{\scriptscriptstyle{\parallel}}{\pm}{2\mu_{0}mH}/{f}=\tilde{q}_{\pm}.
\end{equation}

The explicit form of spinors $\omega^{\pm}_{\zeta}$ is
\begin{equation}
\begin{array}{c}
(q^0+m)\,\omega_{\zeta}^{\pm}=\displaystyle\frac{1}{T}
     \left(
       \begin{array}{c}
         e^{- i \phi/2} \left[ \displaystyle (q^0+m) \mp 2\mu_{0}Hq_{\scriptscriptstyle{\parallel}}/f \right] \\
         e^{i \phi/2}\left[
     (q^0+m)(\pm\zeta R_{\pm}-
    \tilde{q}_{\pm}/q_{\bot}) \mp 2\mu_{0}H {q_\bot}/f
    \right] \\
       \end{array}
     \right),\\[30pt]
{({\bb{\sigma}}\cdot{\bf{q}})}\, \omega_{\zeta}^{\pm}=
\displaystyle\frac{1}{T}\left(
\begin{array}{c}
e^{-i\phi/2}\displaystyle  \left[\pm\zeta R_{\pm}q_{\bot} \mp 2\mu_{0}Hq^{0}/f\right] \\
e^{i\phi/2} \displaystyle \left[q_{\bot}-{q_{\scriptscriptstyle{\parallel}}}(\pm\zeta R_{\pm}-\tilde{q}_{\pm}/{q_{\bot}})
\right] \\
\end{array}
\right),
\end{array}
\end{equation}
\noindent where
\begin{equation}\label{f}
    T_{\pm}=\left\{2R_{\pm}
    \left[
    R_{\pm}\mp{\zeta}
    \left( \frac{\tilde{q}_{\pm}}{q_{\bot}} \pm \displaystyle \frac{2\mu_{0}Hq_{\bot}}{f(q^{0}+m)}
    \right)
    \right]
     \right\}^{1/2}.
\end{equation}

In the selected reference frame the components of canonical momentum
for positive-energy and for negative-energy solutions, respectively,
are
\begin{equation}
    \begin{array}{c}
       \displaystyle P^{1}_{\zeta}(\pm q)=\pm q_{\bot}\cos\phi\left( 1 + \zeta \frac{f \mp 2\mu_{0}Hq_{\scriptscriptstyle{\parallel}}/m}
        {2q_{\bot}R_{\pm}}\right), \\
        \displaystyle P^{2}_{\zeta}(\pm q)=\pm q_{\bot}\sin\phi\left( 1 + \zeta \frac{f \mp 2\mu_{0}Hq_{\scriptscriptstyle{\parallel}}/m}
        {2q_{\bot}R_{\pm}} \right), \\
        \displaystyle P^{3}_{\zeta}(\pm q)=\pm \tilde{q}_{\pm}\left(1+{\zeta}
        \frac{f\mp 2\mu_{0}Hq_{\scriptscriptstyle{\parallel}}/m}{2q_{\bot}R_{\pm}}\right)- 2\mu_{0}mH/f.
    \end{array}
\end{equation}

Consider at first the $\zeta'=\zeta$ case. In this case on the right-hand side of equation (\ref{ort011}) we have
\begin{equation}\label{ort012}
    \delta^{3}({\bf P}_{\zeta}(\pm q')-{\bf P}_{\zeta}(\pm q))= |J_{\zeta}(\pm q)|^{-1}\delta^{3}({\bf q}'-{\bf q}),
\end{equation}
\noindent and as far as ${q'}^2=q^2=m^2$ we get ${q'}^{\mu}={q}^\mu$. Therefore,
\begin{equation}\label{ort014}
 {\mathbb{N}}^{2}_{+} = (2\pi)^{3}
\delta^{3}({\bf q}-{\bf q}').
\end{equation}
\noindent In equation (\ref{ort111}) the delta function
provides the relation
\begin{equation}\label{ort015}
{\bf q}+{\bf q}'=0.
\end{equation}
\noindent So we have
\begin{equation}\label{ort016}
{\mathbb{N}}^{2}_{-}=0.
\end{equation}

Consider next the $\zeta'\neq \zeta$ case.
The delta function provides the following relations:
\begin{equation}\label{rel}
 {\tilde{q}}_{\pm}'/{q'_{\bot}}
 =\tilde{q}_{\pm}/{q_{\bot}}, \quad {\phi}'={\phi},
\end{equation}
\noindent on the right-hand side of equation (\ref{ort011}) and
relations
\begin{equation}\label{rel1}
 {\tilde{q}}_{\pm}'/{q'_{\bot}}
 =-\,\tilde{q}_{\mp}/{q_{\bot}}, \quad {\phi}'={\phi}+\pi,
\end{equation}
\noindent on the right-hand side of equation (\ref{ort111}).

\noindent Using (\ref{rel}) and (\ref{rel1})
and bearing in mind that
$\zeta' = -\zeta$ we find
\begin{equation}\label{}
 \left({\psi^{\pm}_{-\zeta}}(q')\right)^{\dag} \psi^{\pm}_{\zeta}({{q}})=0, \;\left({\psi^{\pm}_{-\zeta}}(q')\right)^{\dag} \psi^{\mp}_{\zeta}({{q}})=0.
\end{equation}

Thus we have proved relations (\ref{ort01}) and (\ref{ort001}), i.e. the orthogonality of system (\ref{Wf}).
It is easy to verify that for the cases where four-vector $f^\mu$ is spacelike or lightlike we can get the same result.


\begin{thebibliography}{99}
\bibitem{x} We use this name because in radiative transitions due to the
self-polarization effect \cite{ST}, the neutral particle obtains a
polarization defined by this vector \cite{L} (see also
\cite{BTB95}).
\end{thebibliography}
\end{document}